\documentclass[aps,prd,12pt,nofootinbib,tightenlines,bibnotes,superscriptaddress,notitlepage]{revtex4-1}

\usepackage{graphicx,float,amsthm,amssymb,verbatim}
\usepackage{hyperref}
\usepackage{caption}

\usepackage{amsfonts}
\usepackage{amsmath,epsfig,bm}
\usepackage{graphicx}
\usepackage{color}

\usepackage{tabularx} 

\usepackage{times}

\newcommand{\be}{\begin{equation}}
\newcommand{\ee}{\end{equation}}
\newcommand{\ba}{\begin{eqnarray}}
\newcommand{\ea}{\end{eqnarray}}
\newcommand{\bd}{\begin{displaymath}}
\newcommand{\ed}{\end{displaymath}}

\renewcommand{\vec}[1]{\mbox{\boldmath$#1$}}
\def\thalf{{\textstyle{\frac{1}{2}}}}

\def\l{\left}
\def\r{\right}

\begin{document}

\title{Calculating Fluctuations and Self-Correlations Numerically for Causal Charge Diffusion in Relativistic Heavy-Ion Collisions}
\author{Aritra De}
\affiliation{School of Physics \& Astronomy, University of Minnesota, Minneapolis, MN 55455, USA}
\author{Christopher Plumberg}
\affiliation{Department of Astronomy and Theoretical Physics, Lund University, Sölvegatan 14A, SE-223 62 Lund, Sweden}
\author{Joseph I. Kapusta}
\affiliation{School of Physics \& Astronomy, University of Minnesota, Minneapolis, MN 55455, USA}

\vspace{.3cm}
\date{\today}

\parindent=20pt

\begin{abstract}
We study the propagation and diffusion of electric charge fluctuations in the Bjorken hydrodynamic model with both white and Catteneo noise using purely numerical methods. We show that a global lattice of noise fluctuations is required to fully calculate the two-point correlators of charge.  We solve the stochastic differential equations that arise from the charge conservation equation on the lattice. We explicitly identify the self-correlation term in the case of Catteneo noise and provide a physical interpretation.  We provide a numerical recipe to remove this contribution from the full two-point correlators.  Finally, we calculate the balance functions for charged hadrons. By limiting the speed of signal propagation, we observe the expected narrowing of the balance functions after removing the self-correlations.
\end{abstract}

\maketitle

\section{Introduction}

Relativistic hydrodynamics is used to study not only the equation of state but also dynamical quantities, such as the transport coefficients, of the quark-gluon plasma. The applicability of hydrodynamics is justified if the mean free paths of the particles are small compared to the distances over which thermodynamic quantities vary. It turns out that hydrodynamics is very successful in modeling high energy nuclear collisions. There are experimental facilities which produce and study quark-gluon plasma: the Relativistic Heavy Ion Collider (RHIC) at Brookhaven National Laboratory and the Large Hadron Collider (LHC) at CERN. Fluid equations describe conservation of energy, momentum, baryon number, electric charge, and strangeness. Anisotropic particle production, such as elliptic flow, in heavy ion collisions gives credence to the use of hydrodynamics in simulating these collisions. It has been successful in describing various properties like particle spectra, particle correlations, and in obtaining values of transport quantities like the ratio of shear viscosity to entropy density $\eta/s$. By comparing particle spectra with experimental data, hydrodynamical simulations also help in understanding the initial state, its fluctuations, and hence properties of strongly interacting matter in general. Initially, the assumption of ideal hydrodynamics worked very well in describing the data which indicated that the system was strongly interacting. Later, important aspects like the lattice-computed QCD equation of state and viscous properties were taken into account to study transport properties of quark-gluon plasma with more precision.

The fluctuation-dissipation theorem relates the dissipative properties of a system to its hydrodynamical fluctuations. In particular, it allows us to infer quantities like shear and bulk viscosity, and electrical conductivity, from the magnitude of fluctuations. Hydrodynamical fluctuations have also been used to study static critical phenomena \cite{torres} near a possible critical point. More recently, there have been studies of dynamic critical phenomena near a QCD critical point \cite{teaney,yiyin,nahrgang,Bluhm}. Critical points are characterized by large fluctuations. This led to the suggestion to study fluctuations in conserved quantities, such as electric charge, baryon number, and strangeness on an event-by-event basis \cite{stephanov_shuryak}.  It has also been suggested to study non-gaussianities (higher order cumulants) of the fluctuations near critical points as they are more sensitive to large correlation lengths \cite{stephanovprl}. Correlation lengths theoretically diverge near the critical points, but in the scenario of the heavy ion collisions are limited by the finite system size \cite{stephanov_shuryak} as well as by the finite lifetime of the system \cite{berdnikov}. 

Thus it becomes imperative to study the hydrodynamics of fluctuations in the context of heavy ion collisions. The relativistic theory of hydrodynamic fluctuations in the context of heavy-ion collisions was introduced in Ref. \cite{kapusta2012}.  In the current work, we focus on calculating the two-point correlations of charge fluctuations and the resulting balance functions of pions. The balance function measures the difference in probability of finding a particle of opposite charge in another fluid cell versus a particle of same charge given a charged particle in a given fluid cell \cite{spratt}.  This problem was studied analytically in the 1+1 dimensional Bjorken model in Ref. \cite{plumbergkapusta}.  Analytic calculations are not possible for state of the art 3+1 dimensional, non-boost invariant, hydrodynamics.  In preparation for extensions to modern hydrodynamic models, we develop numerical methods to solve the relevant stochastic differential equations numerically.  Particular attention is paid to the physical interpretation of self-correlations and how they can be subtracted to make comparison to experimental data.

The outline of the paper is as follows. In Sec. \ref{noise} we review the normal diffusion, the Cattaneo, and the Gurtin-Pipkin equations, and discuss how self-correlations arise.  In Sec. \ref{1+1} we outline the application of the relevant stochastic differential equations in the context of the Bjorken hydrodynamic model for heavy-ion collisions.  In Sec. \ref{solving} we present solutions to those equations.  In Sec. \ref{selfies} we show how self-correlations can be clearly identified. In Sec. \ref{balance} we calculate the balance functions that relate theory to experiment.  Our conclusions are presented in Sec. \ref{conclude}.  Details of how the stochastic differential equations are solved are presented in the Appendices.  The numerical method is readily transferrable to heavy-ion collisions which have no spatial symmetries and thus useful for future calculations.

\section{Noise, Fluctuations and Self-Correlations}
\label{noise}

Since the usual diffusion equation leads to instantaneous signal propagation, which is inconsistent with special relativity, one needs a diffusion equation which is the same order in spatial and temporal derivatives with characteristic relaxation times and lengths. In this paper we solve the simplest diffusion equation satisfying this condition, called the Catteneo equation \cite{catteneo}, numerically. The resulting differential equation is a stochastic differential equation (SDE) because it contains random noise terms. The way to solve SDEs is to solve the differential equation for a large number of events (here on the order of 1 million or more) and study the correlation functions.  A finite difference method is used to solve the SDE.

Consider the ordinary diffusion equation with white noise.  In the context of the Bjorken model, which has boost invariance and no dependence on transverse coordinates, the two variables are proper time $\tau = \sqrt{t^2 - z^2}$ and spatial rapidity $\xi = \thalf \ln[(t+z)/(t-z)]$, where the beam axis is along the $z$ direction.   The noise $f$, appropriately defined (see below), is a dimensionless random variable with correlator 
\be
\langle f(\tau_1, \xi_1) f(\tau_2, \xi_2) \rangle =  \frac{N(\tau_2)}{2\pi} \delta(\tau_1-\tau_2)\delta(\xi_1-\xi_2) \,,
\ee
which is a product of Dirac $\delta$-functions in time and space with normalization determined by the fluctuation-dissipation theorem
\be
N(\tau) = \frac{4 \pi \sigma_Q(\tau) T(\tau)}{A \tau s^2(\tau)} \,.
\ee
Here $\sigma_Q$ is the charge conductivity, $T$ is the temperature, $s$ is the entropy density, and $A$ is the transverse area. To generate this numerically on a discrete lattice with spacings $\Delta \xi$ and $\Delta \tau$, we sample $f$ from a normal distribution with zero mean and variance 
$N(\tau)/(2\pi\Delta \xi \Delta \tau)$. The analysis of how finite difference methods work computationally for solving SDEs is discussed in Appendix A.

Consider the difference between white and colored noise. The standard two-point function for white noise in frequency and momentum space is
\ba
\langle \tilde{f}(\omega_1, k_1) \tilde{f}(\omega_2, k_2) \rangle 
&=& \int d\tau_1 \, d\tau_2 \, d\xi_1 \, d\xi_2 
\, e^{-i(k_1\xi_1 + k_2\xi_2)} \, e^{-i(\omega_1\tau_1 + \omega_2\tau_2)} \langle f(\tau_1, \xi_1) f(\tau_2, \xi_2) \rangle \nonumber \\
&=& \delta(k_1+k_2)  \tilde{N}(\omega_1+\omega_2)
\ea
where $\tilde{N} $ is the Fourier transform of $N$.  Generalizing this to Catteneo noise (which is an example of colored noise), we recall that the two-point function for the noise obeys \cite{kapustayoung}. 
\be
\langle (\tau_Q \, \partial \tau_1 +1)\tilde{f}(k_1,\tau_1) (\tau_Q \, \partial \tau_2 +1)\tilde{f}(k_2,\tau_2)\rangle
 = N(\tau_1) \delta(\tau_1-\tau_2)\delta(k_1 + k_2) \,
\ee
where $\tau_Q$ is a relaxation time. In frequency and momentum space this becomes
\be
\langle \tilde{f}(\omega_1, k_1) \tilde{f}(\omega_2, k_2) \rangle = \frac{\delta(k_1+k_2)  
\tilde{N}(\omega_1+\omega_2)}{(i\tau_Q \omega_1+1)(i\tau_Q \omega_2+1)} \,.
\ee
The noise correlator is no longer a Dirac $\delta$-function in time anymore; instead, it is smeared out, hence the name colored noise.

\begin{figure}[H]
\centering
\includegraphics[scale=0.42]{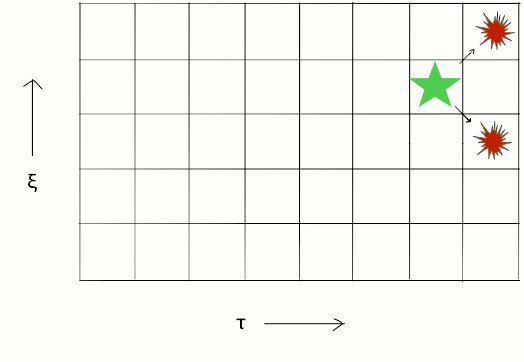}
\caption{An example of either white or colored noise.  A fluctuation in one cell, represented by a star, causes a correlation between two cells in the next time step, represented by bursts, separated in space from each other and from the original fluctuation. (color online)}
\label{color1}
\end{figure}

\begin{figure}[H]
\centering
\includegraphics[scale=0.42]{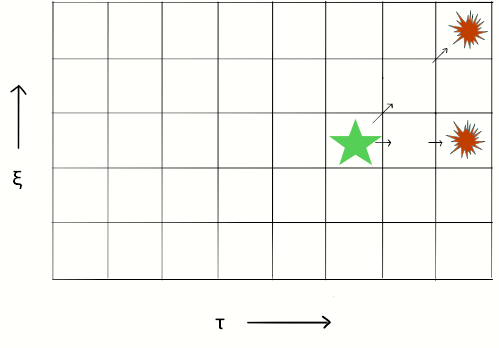}
\caption{An example of either white or colored noise.  A fluctuation in one cell, represented by a star, causes a correlation between two cells two time steps later, represented by bursts, but only one is separated in space from the original fluctuation. (color online)}
\label{color2}
\end{figure}

\begin{figure}[H]
\centering
\includegraphics[scale=0.42]{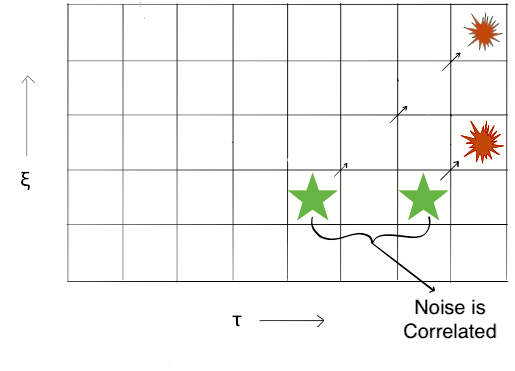}
\caption{An example of colored noise.  Fluctuations at the same point in space but at different times are correlated, as represented by the stars.  This results in a correlation between the two cells, represented by bursts. (color online)}
\label{color3}
\end{figure}

The following three figures will help illustrate some of the physics to come.  Figure \ref{color1} shows a fluctuation, represented by a star, in a particular spacetime cell.  The signal, represented by bursts, is transmitted to the two adjacent spatial cells in the next time step.  Hence those two cells have correlated fluctuations.  This type of correlation can arise from either white or colored noise.  Figure \ref{color2} shows a fluctuation in one spacetime cell with its signal transmitted to two spacetime cells two time steps later.  This type of correlation can also happen with either white or colored noise.  Figure \ref{color3} shows a situation that only happens with colored noise.  The two stars are correlated, and their signals lead to correlations between the same two cells as shown in the previous figures.

Self-correlations arise from correlations in the same spatial cell.  For white noise this means the star and the burst are in the same spacetime cell.  In discretized spacetime this leads to a Kronecker $\delta$-function in $\xi$, while in the continuum limit this leads to a Dirac $\delta$-function.  The latter is somewhat unphysical, since all correlations have some finite extent.  For colored noise, the self-correlation begins in the cell hosting the original fluctuation, and then continues in subsequent time steps but always in the same spatial cell due to the time-correlated nature of colored noise. Noise generated at a previous time in the same spatial cell will hydrodynamically evolve to a correlated charge fluctuation in a different spatial fluid cell. Hence the self-correlation will be non-trivial for colored noise.

\begin{figure}[H]
\centering
\includegraphics[scale=0.5]{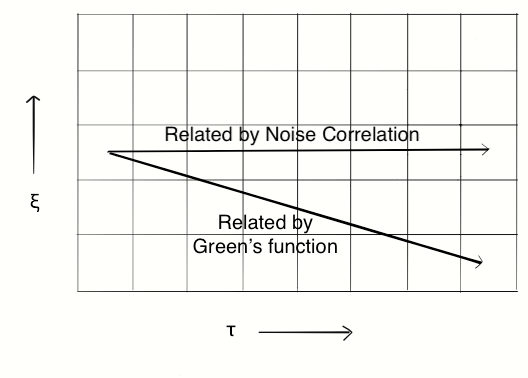}
\caption{Schematic of the lattice setup for Catteneo noise.}
\label{prop}
\end{figure}

Figure \ref{prop} shows another way to visualize the colored Cattaneo noise.  At a fixed spatial cell, correlations arise at different times due to $\tau_Q > 0$.  Correlations also propagate to other spatial cells with increasing time via a Green's function.  The mathematical formalism and details of how it is implemented numerically with be presented in the following sections.

The final charge correlations are determined at some $\tau_f$.  One must integrate over all prior times $\tau_i \le \tau \le \tau_f$ to obtain the final time charge correlators. Thus one can define self-correlations as the correlation of a charge fluctuation generated in $\xi_1$ at final time $\tau_f$ with another charge fluctuation generated at the same $\xi_1$ but at a previous time and hence had time to travel to a different $\xi_2$ at $\tau_f$. It is non-trivial for colored noise because colored noise generated in same $\xi$ are correlated in time.

One can go further and consider the Gurtin-Pipkin noise \cite{gurtin} which introduces a noise correlation in spatial rapidity in addition to the correlation in proper time. Gurtin-Pipkin noise has been dealt with analytically in Ref. \cite{kapustayoung}.  In Cartesian coordinates Gurtin-Pipkin noise results in the following diffusion equation
\be
\left[ \frac{\partial}{\partial t} - D_Q \nabla^2 + \tau_Q \frac{\partial^2}{\partial t^2} + \tau_2^2 \frac{\partial^3}{\partial t^3} - \tau_3 D_Q \frac{\partial}{\partial t} \nabla^2 \right] n_Q = 0 \,.
\label{GPeq}
\ee
Numerical simulation of Gurtin-Pipkin noise is deferred to a future work.

\section{Diffusion in Boost Invariant 1+1 Hydrodynamics}
\label{1+1}

This section is a mini-review of the problem addressed previously in Ref. \cite{plumbergkapusta} to help setup the use of numerical methods for solving the resulting SDE.  We will work in 1+1 dimensional boost-invariant Bjorken hydrodynamics. The longitudinal boost-invariance implies that the initial conditions for local variables are only functions of the proper time $\tau$.  We neglect the bulk and shear viscosities in order to focus on charge transport. 

The energy-momentum tensor for an ideal fluid is
\be
T^{\mu \nu} = w u^{\mu}u^{\nu} - pg^{\mu \nu} \,.
\ee 
We take the Landau-Lifshitz approach where $u^{\mu}$ is the velocity of energy transport. The electric current takes the form 
\be
J_Q^{\mu} = n_Q u^{\mu} + \Delta J^{\mu}
\ee
where $n_Q$ is the proper charge density and $\Delta J^{\mu}$ is the dissipative part.  In first-order viscous fluid dynamics $\Delta J^{\mu}$ takes the form 
\be
\Delta J^{\mu} =  D_Q \Delta^{\mu}n_Q = \sigma_Q \Delta^{\mu} \mu_Q
\ee
where $\mu_Q$ is the charge chemical potential, $\sigma_Q$ is the charge conductivity and $\Delta^{\mu}$ is the transverse derivative
\be
\Delta^{\mu} = \partial^{\mu} - u^{\mu}(u\cdot \partial) \,.
\ee
Conventional charge diffusion follows the usual diffusion equation
\be
\left( \frac{\partial }{\partial t} -D_Q \nabla^2  \right) n_Q =0  \,.
\ee
The diffusion constant $D_Q$ and charge conductivity are related by the Einstein relation $D_Q = \sigma_Q/\chi_Q$, where $\chi_Q$ is the electric charge susceptibility defined by
\be
\chi_Q = \frac{\partial n_Q(T,\mu_Q)}{\partial \mu_Q} \,.
\ee
The diffusion equation leads to an infinite speed of propagation which is unphysical and not suitable for hydrodynamic simulations of heavy-ion collision. Therefore the usual diffusion equation is replaced by one with a double derivative in time with a relaxation time factor $\tau_Q$.
\be
\left( \frac{\partial }{\partial t} -D_Q \nabla^2  + \tau_Q \frac{\partial^2}{\partial t^2} \right) n_Q =0  
\ee
This equation is called the Cattaneo equation\hspace{3pt} \cite{catteneo}. It is a combination of the diffusion equation with the wave equation. The dissipative current gets modified to
\be
\Delta J^{\mu} = D_Q \Delta^{\mu}\left[ \frac{1}{1+\tau_Q(u\cdot \partial )}\right] n_Q 
\ee
One can show that high frequency waves travel at a speed of $v_Q = \sqrt{D_Q/\tau_Q}$ \cite{kapustayoung}.
The fluctuation-dissipation theorem relates the two-point function, which provides a measure of the variance of fluctuations, to the dissipation from diffusion. A stochastic noise term $I^{\mu}$ is therefore added to the charge current. 
\be
J^{\mu}  =  n_Qu^{\mu} + \Delta J^{\mu} + I^{\mu} 
\ee
One-point functions vanish and the two-point functions are determined by the fluctuation-dissipation theorem. For the usual diffusion equation
\be
\langle I^{\mu}(x)\rangle =0 \qquad \langle I^{\mu}(x_1)I^{\nu}(x_2) \rangle = 2 \sigma_Q T \, h^{\mu\nu} \delta(x_1-x_2)
\ee
where $h^{\mu\nu} = u^{\mu}u^{\nu} - g^{\mu\nu}$ is the transverse projector. This is white noise. In the Catteneo equation the fluctuations are
\be
\langle I^{i}(x_1)I^{j}(x_2) \rangle =  \frac{\sigma_Q T}{\tau_Q}  \delta(\vec{x}_1-\vec{ x}_2) \, e^{-|t_1-t_2|/\tau_Q} \, \delta_{ij}
\ee
The delta function in time is replaced by an exponential decay function.  In the limit $\tau_Q \rightarrow 0$ this two-point function becomes the Dirac $\delta$- function for white noise. 

The following are the relations between the Cartesian coordinates and the proper time and spatial rapidity appropriate for the Bjorken model.
\begin{eqnarray}
\begin{aligned}
t &=& \tau \cosh \xi \qquad z &= \tau \sinh \xi \\
\tau &=& \sqrt{t^2-z^2} \qquad \xi &= \tanh^{-1}\left( \frac{z}{t} \right) \\
\end{aligned}
\end{eqnarray}
The flow velocity is
\be
u^0 = \cosh \xi \quad u^z = \sinh \xi \,.
\ee
The transverse derivatives are
\begin{eqnarray}
\Delta^0 = -\frac{\sinh \xi}{\tau} \frac{\partial }{\partial \xi} \qquad \Delta^3 = -\frac{\cosh \xi}{\tau} \frac{\partial }{\partial \xi} \quad  \text{with} \quad u\cdot \partial = \frac{\partial}{\partial \tau} \,.
\end{eqnarray}
The fluctuating contribution to the current is written as
\begin{eqnarray}
I^0 &=& s(\tau) f(\xi,\tau) \sinh \xi \\
I^3 &=& s(\tau) f(\xi,\tau) \cosh \xi \,.
\end{eqnarray}
The entropy density $s$ is factored out to make $f$ dimensionless.  The background fluid equations for the proper charge density and entropy density are 
\begin{eqnarray}
\frac{d s}{d \tau} + \frac{s}{\tau} = 0 \;\; &\Rightarrow& \;\; s(\tau) = \frac{s_i \tau_i}{\tau} \\
\frac{d n_Q}{d \tau} + \frac{n_Q}{\tau} = 0 \;\; &\Rightarrow& \;\; n_Q(\tau) = \frac{n_i \tau_i}{\tau} \,.
\end{eqnarray}
These are a manifestation of the conservation of entropy and charge, respectively. The $s_i$ and $n_i$ are the densities at some initial time $\tau_i$. We take the initial proper charge density $n_i$ to be zero, hence the average charge density for subsequent times is zero as well. 

Now let us look at the charge current conservation equation $\partial_{\mu}J^{\mu} = 0$.  
It is convenient to define the variable $X = \tau \delta n$ because, in the absence of fluctuations, this quantity is conserved during the hydrodynamic evolution.  After a few steps of algebra the full charge conservation equation becomes
\bd
\left[ \frac{\tau}{D_Q\chi_Q T} + \tau_Q \frac{\partial}{\partial \tau} \left(\frac{\tau}{D_Q\chi_Q T}\right) \right] \frac{\partial X}{\partial \tau}
+ \frac{\tau_Q \tau }{D_Q\chi_Q T} \frac{\partial^2 X}{\partial \tau^2}  - \frac{1}{\tau \chi_Q T} \frac{\partial^2 X}{\partial \xi^2}
\ed
\be 
+ \left[ \frac{\tau s}{D_Q\chi_Q T}+ \tau_Q \frac{\partial}{\partial \tau}\left(\frac{\tau s}{D_Q\chi_Q T}\right)\right]\frac{\partial f}{\partial \xi} + 
\frac{\tau_Q \tau s}{D_Q\chi_Q T}\frac{\partial^2 f}{\partial \xi \partial \tau} = 0 \,.
\label{diffeqcolor}
\ee
For the case $\tau_Q=0$ (usual diffusion equation) this simplifies to
\be 
\frac{\partial X}{\partial \tau} - \frac{D_Q}{\tau^2}\frac{\partial^2 X}{\partial \xi^2} + s\frac{\partial f}{\partial \xi} = 0 \,.
\ee
Due to boost invariance it is useful to use the Fourier transform 
\be
X(\xi,\tau) = \int_{-\infty}^{\infty} \frac{dk}{2\pi } e^{ik\xi} \tilde{X}(k,\tau) \,,
\ee
and similarly for $f$.  Then the SDE for white noise is
\be
\frac{\partial }{\partial \tau} \tilde{X} + \frac{D_Q k^2}{\tau^2} \tilde{X} = -iks\tilde{f}
\ee
and for colored Cattaneo noise
\bd
\left[ \frac{\tau}{D_Q\chi_Q T} + \tau_Q \frac{\partial}{\partial \tau} \left(\frac{\tau}{D_Q\chi_Q T}\right) \right] \frac{\partial \tilde{X}}{\partial \tau}
+ \frac{\tau_Q \tau }{D_Q\chi_Q T} \frac{\partial^2 \tilde{X}}{\partial \tau^2}  + \frac{k^2}{\tau \chi_Q T} \tilde{X}
\ed
\be 
= -ik \left[ \frac{\tau s}{D_Q\chi_Q T}+ \tau_Q \frac{\partial}{\partial \tau}\left(\frac{\tau s}{D_Q\chi_Q T}\right)\right] \tilde{f} -i 
\frac{k \tau_Q \tau s}{D_Q\chi_Q T}\frac{\partial \tilde{f}}{\partial \tau} \,.
\label{color-kspace}
\ee

For the sake of comparison and for definiteness, we follow Ref. \cite{plumbergkapusta} and assume both $D_Q$ and $\tau_Q$ are constant within the range of temperature to be considered.  This means that high frequency waves propagate with a constant value of $v_Q$.  For the same reasons we assume that 
$s \sim T^3$ and $\chi \sim T^2$.  Hence $T \sim \tau^{-1/3}$.

\section{Solving the Stochastic Differential Equations}
\label{solving}
We start by solving the stochastic differential equation for white noise. As explained earlier, we will solve it on a spacetime lattice and choose spacings $\Delta \xi = 0.09$ and $\Delta \tau = 10^{-4}$ fm/c. We set the parameters such that $( \chi(\tau_f) T_f)/(\tau_f \Delta \xi) = 0.5122$ MeV$^3$ fm$^{-3}$. We source the noise function $f$ from a normal distribution with mean $0$ and variance $1/\sqrt{\Delta t \Delta \xi}$. The density-density correlator arising from the noise fluctuation which is a solution to the SDE in our discretized system, evaluted at the final time $\tau_f$, has the analytical form
\be
\langle \delta n (\xi_1, \tau_f) \delta n (\xi_2, \tau_f) \rangle 
= \frac{\chi_Q(\tau_f)T_f}{A \tau_f} \left[ \frac{\delta_{\xi_1 , \xi_2}}{\Delta \xi}  - \frac{1}{\sqrt{\pi w^2}} e^{-(\xi_1 - \xi_2 )^2/w^2} \right]
\label{whitenn} 
\ee
where
\be
w^2  = 8D_Q \left(\frac{1}{\tau_i} - \frac{1}{\tau_f}\right) \,.
\ee
In the continuum limit $\delta_{\xi_1 , \xi_2}/\Delta \xi \rightarrow \delta(\xi_1 - \xi_2)$.  The parameters chosen for this work are the same as in Ref. \cite{plumbergkapusta}, namely $\tau_i = 0.5$ fm/c, $\tau_f = 6.352$ fm/c, $T_i = 350$ MeV, and $T_f = 150$ MeV. 
We use diffusion constant $D_Q = 0.162 \:\text{fm}$ which is an average over the temperature interval from 150 to 350 MeV taken from Ref. \cite{Aarts}. The equation of state used is the same as in Ref. \cite{torres}, which is $\chi_Q =  \frac{2}{3} T^2$ (including up, down and strange quarks).

The details of how we solve an SDE are discussed in Appendix A.  The solution is presented in Fig. \ref{white2compare}.  The dots represent the result of the SDE simulation for ten million random events.  The solid curve is the Gaussian from Eq. (\ref{whitenn}); it overlays the dots within the width of the line. The Kronecker $\delta$-function at $\xi = 0$ is clearly evident.  
\begin{figure}
\centering
\includegraphics[scale=0.13]{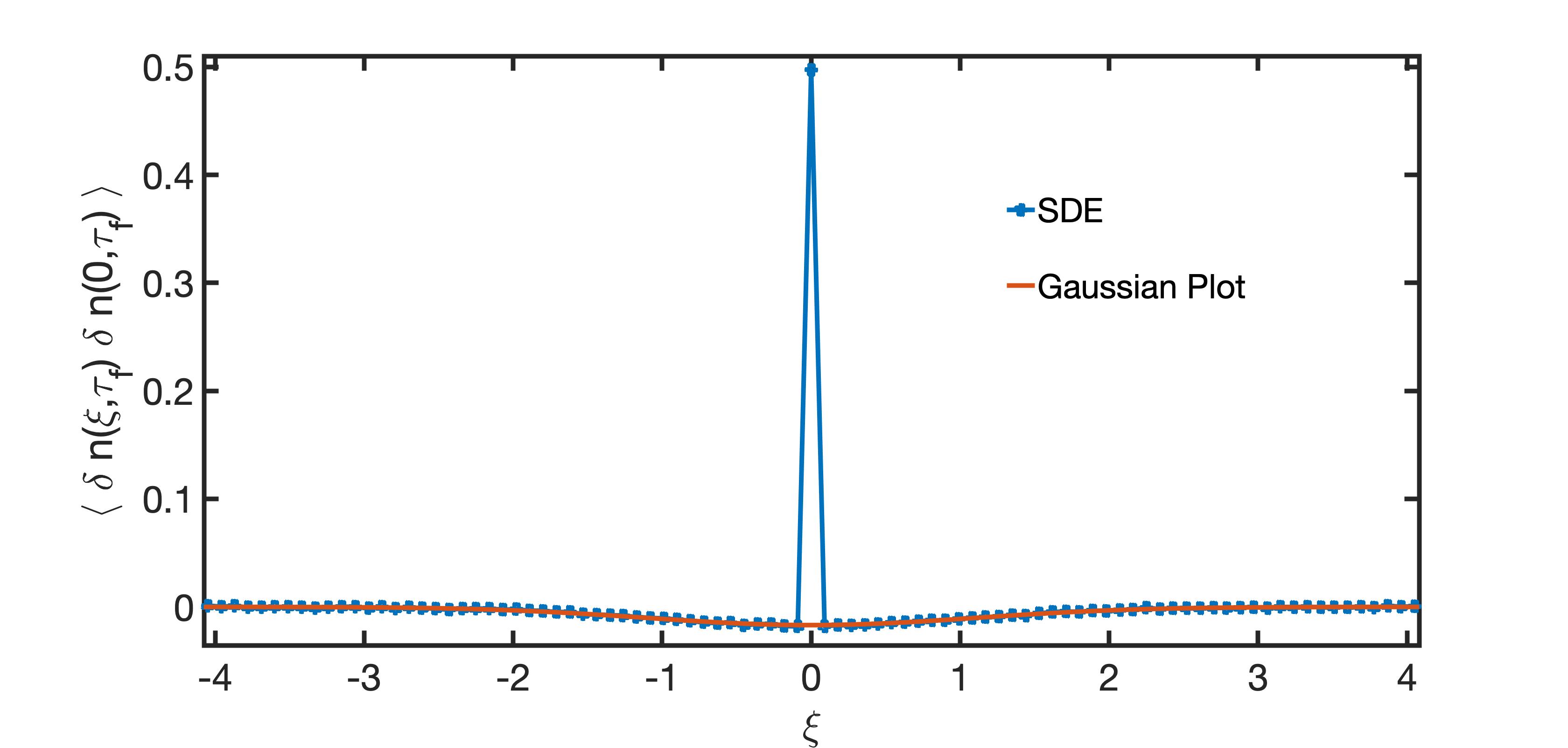}
\caption{White noise density-density correlation function for 10 million events.  The solid curve is the Gaussian from Eq. \ref{whitenn}. (color online)}
\label{white2compare}
\end{figure}

\begin{figure}
\centering
\includegraphics[scale=0.13]{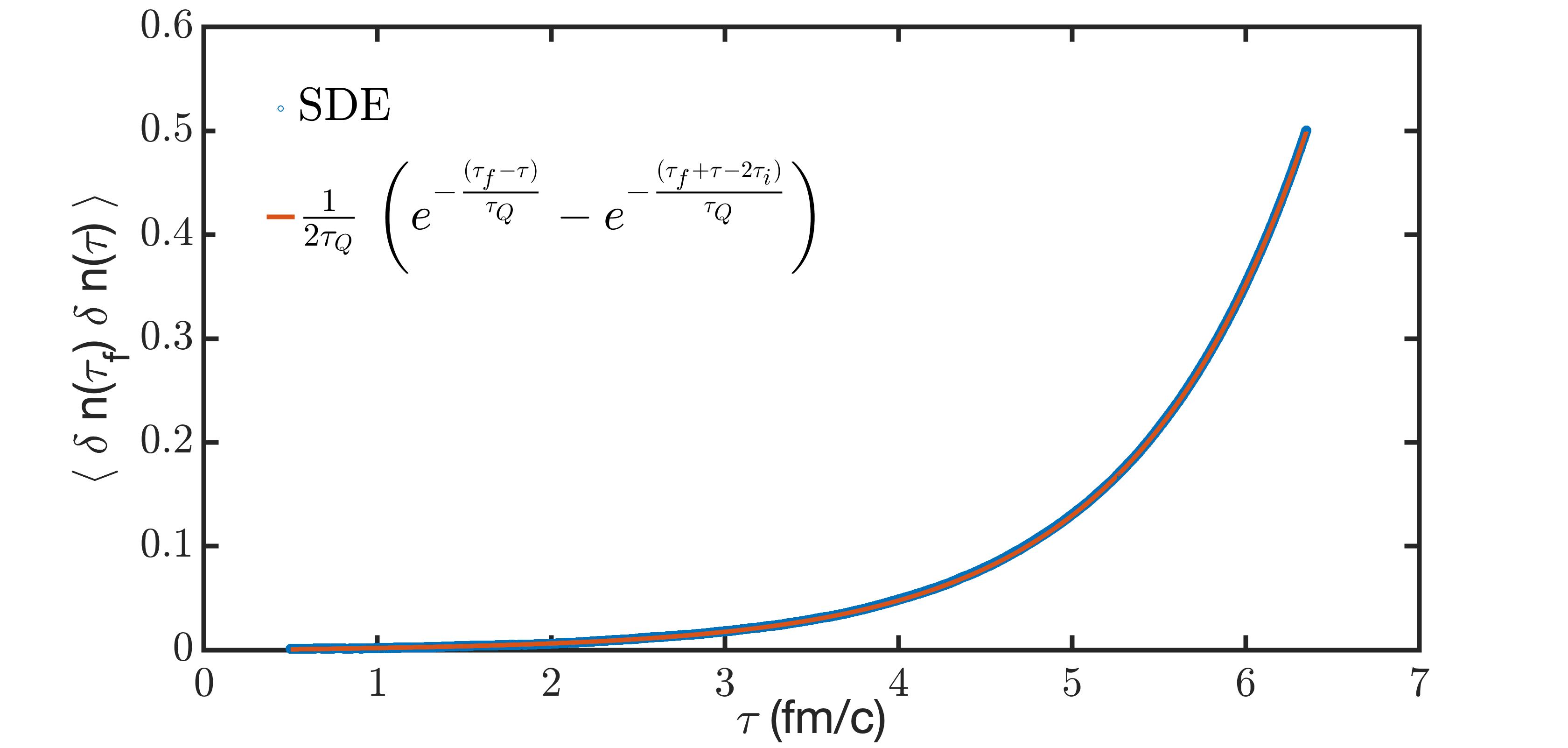}
\caption{Comparison of numerical and analytical results for $v_Q^2 = 0.16$ with rapidity dependences suppressed. (color online)}
\label{exp}
\end{figure}

Next we turn to colored noise.  We have to generate a noise that has the desired correlation in proper time but is uncorrelated in rapidity. The way we do that is by solving another SDE which is called the Langevin equation.
\begin{equation}
f + \tau_Q \frac{\partial f}{\partial \tau} = \zeta
\end{equation}
Here $\zeta$ is the regular white noise.  The relaxation time $\tau_Q$ smoothens the Dirac $\delta$-correlation in proper time. The $\tau_Q$ also introduces the maximum mode velocity to be $v_Q^2 = D_Q/\tau_Q$, thereby removing instantaneous signal propagation. The analytical solution to the Langevin equation (with rapidity dependences suppressed) is
\be
\langle f(\tau_1)f(\tau_2) \rangle = \frac{N(\tau_2)}{4\pi \tau_Q} \left[ e^{-|\tau_1-\tau_2|/\tau_Q} - e^{(2\tau_i-\tau_1-\tau_2)/\tau_Q} \right] 
\equiv {\cal N}(\tau_1,\tau_2) \,.
\label{2pt}
\ee
The derivation is given in Appendix B.  The numerically computed two-point function is plotted in Fig. \ref{exp}. The expected result (\ref{2pt}) and the numerical result are consistent for ten million simulated events.
The grid sizes chosen ensures that they obey the Courant Friedrichs Lewy (CFL) condition \cite{CFL}. This condition states that the numerical domain of dependence of any point in space and time must include the analytical domain of dependence.  Physically, this condition amounts to a signal propagating no more than one spatial cell away during one time step.  For speed $v_Q$ being a constant, this amounts to the condition $\Delta \tau /\tau  < \Delta \xi /v_Q$. 

\begin{figure}[H]
\centering
\includegraphics[scale=0.13]{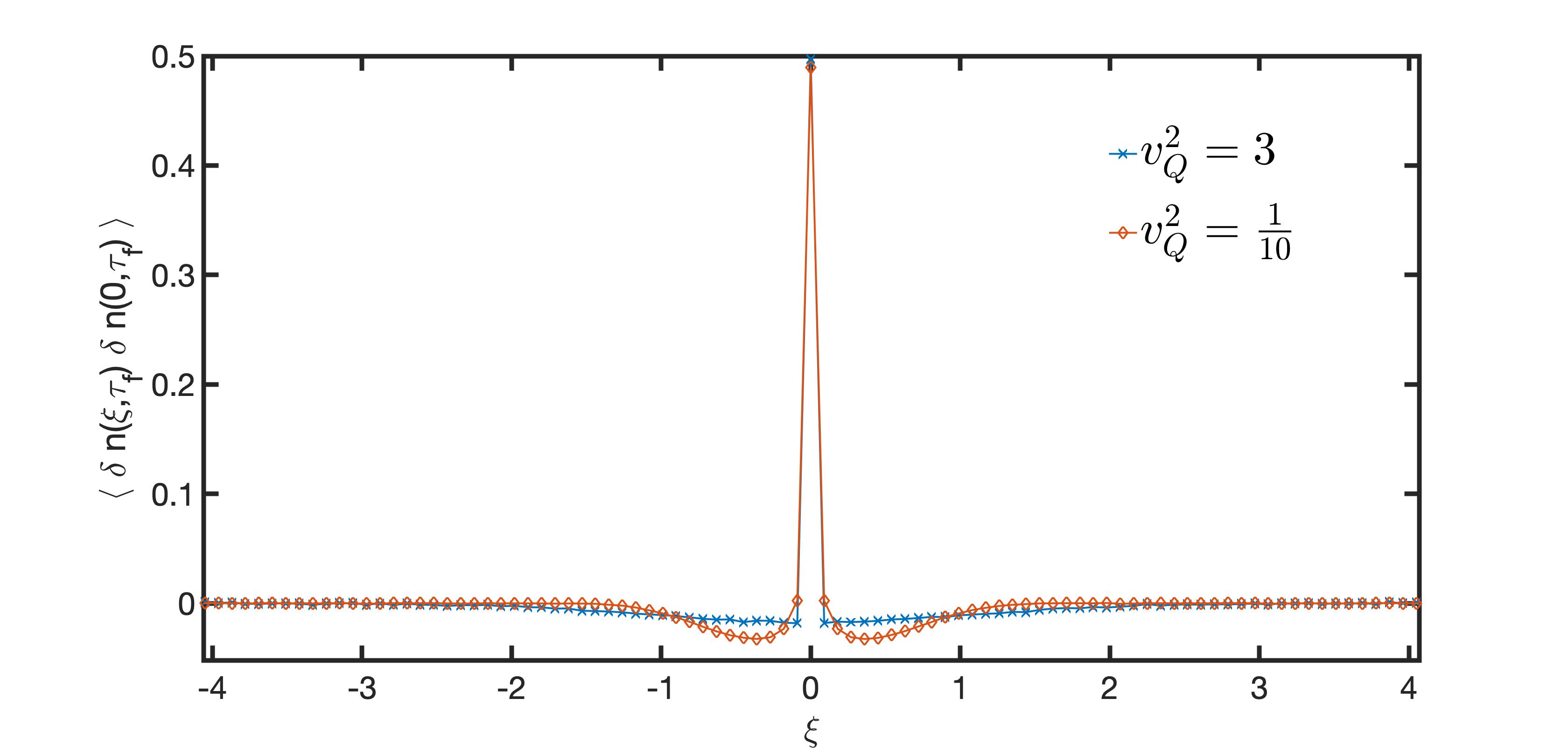}
\caption{Variation of the density-density correlator with the propogation speed $v^2_Q$. (color online)}
\label{2tau}
\end{figure}

Figure \ref{2tau} shows the dependence of the two-point correlator for two very different values of the propogation speed, or equivalently the relaxation time $\tau_Q$.
 
\section{Characterizing Self-Correlations}
\label{selfies}

The self-correlations are trivial for the case of white noise; it's a Dirac $\delta$-function. Even the two-point correlation function of a free Boltzmann gas has a $\delta$- function term \cite{landau1,landau2}.
\be
\langle \delta n(\textbf{x}_1) \delta n(\textbf{x}_2) \rangle = \chi T \delta^3(\textbf{x}_1-\textbf{x}_2) + \cdots   
\ee
This is explained in the Ref. \cite{landau2} where $\langle (\Delta N)^2 \rangle  = \chi T V$. One can see this in Eq. (\ref{whitenn}), where the denomintor $\tau A$ is the Jacobian factor from the Bjorken expansion instead of stationary Cartesian coordinates.  Experiments measure just the two-particle correlation and hence we have to subtract the self-correlation \cite{spratt}. The challenge is to characterize the self-correlations in the presence of colored noise, since it is no longer a Dirac 
$\delta$-function.

\begin{figure}[H]
\centering
\includegraphics[scale=0.12]{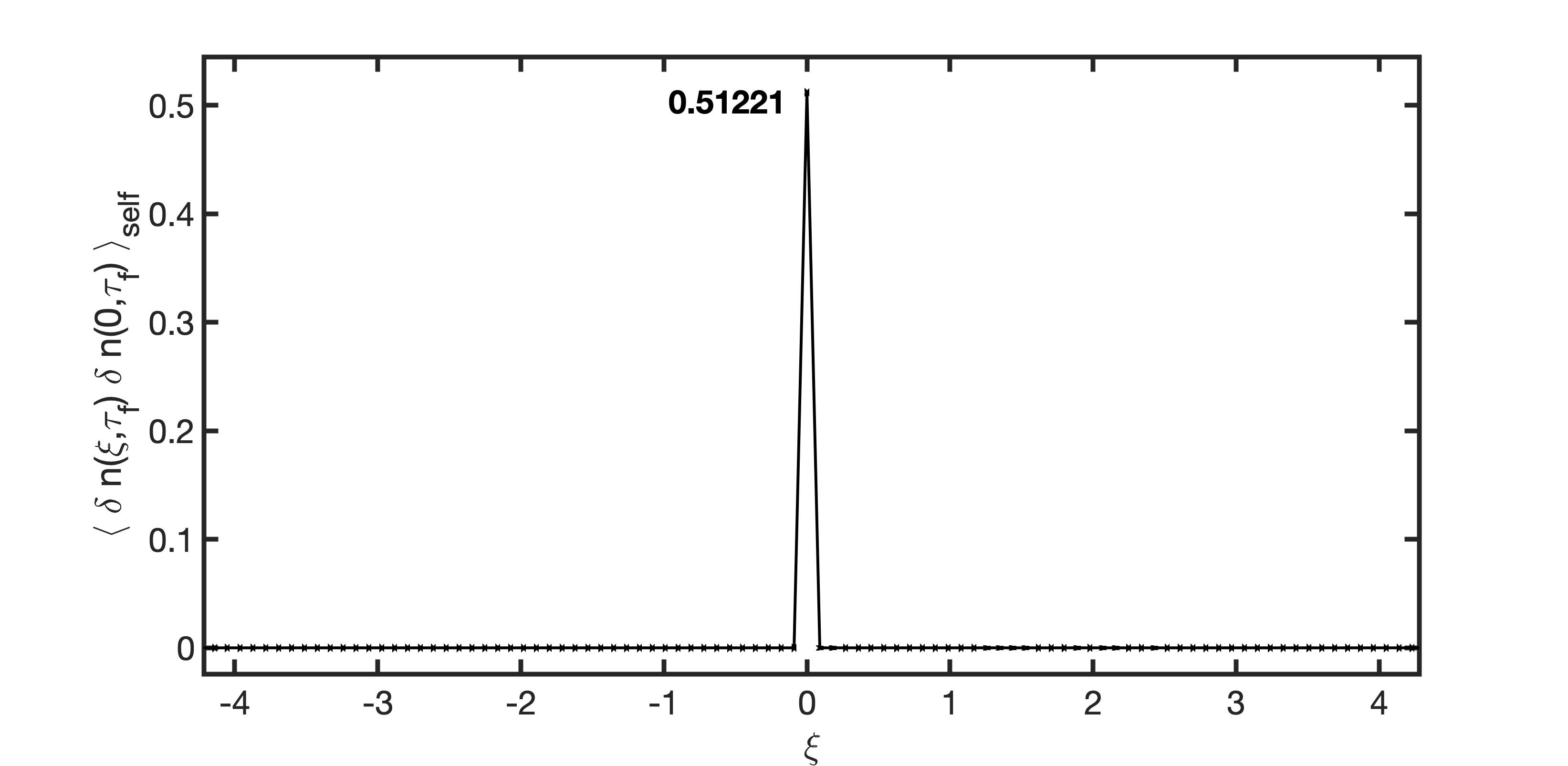}
\caption{The self-correlation.}
\label{self-white}
\end{figure}

\begin{figure}[H]
\centering
\includegraphics[scale=0.12]{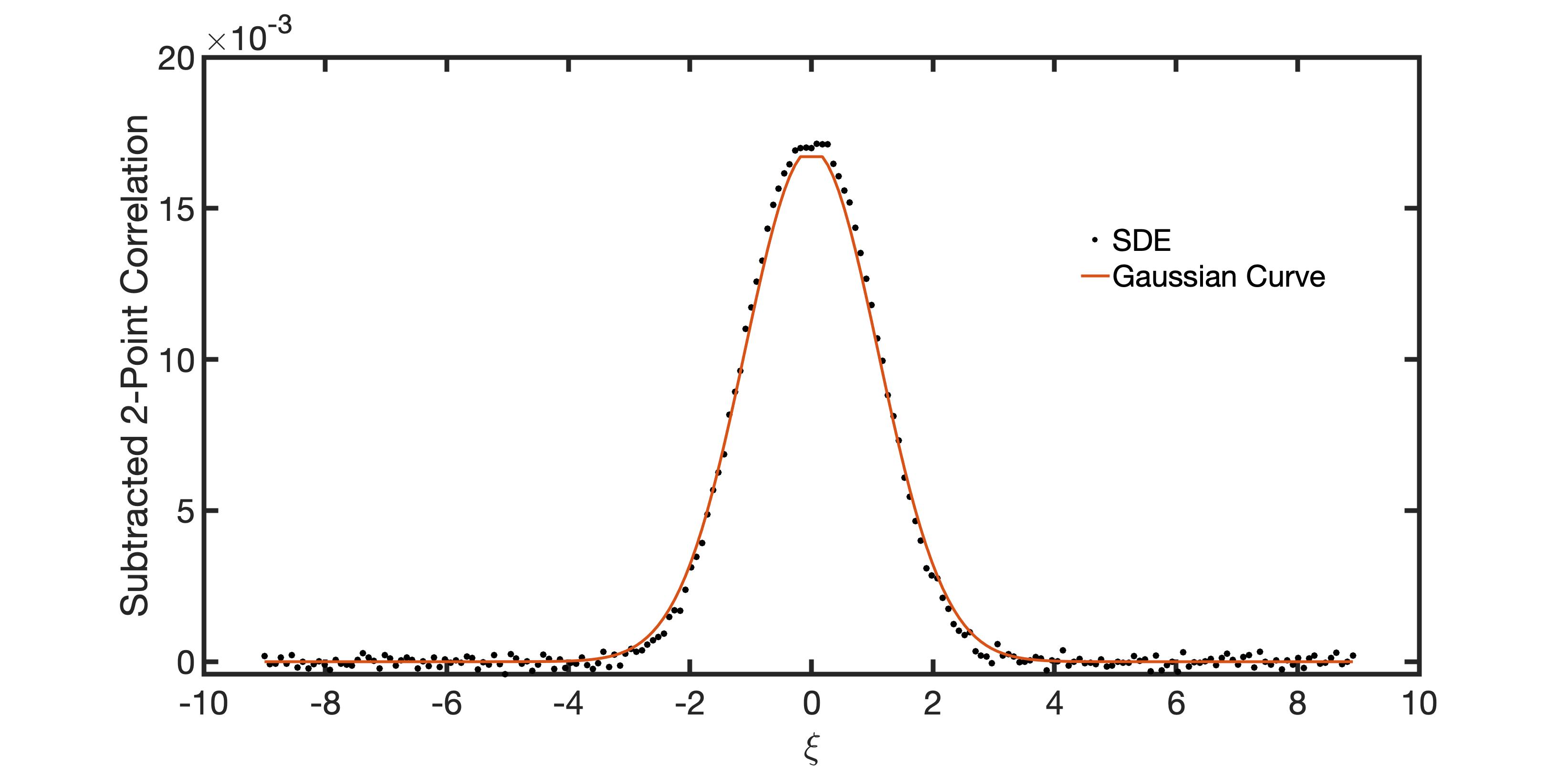}
\caption{The solid curve is the expected Gaussian while the dots represent the result of the SDE simulation for 1 million random events. (color online)}
\label{white-sub}
\end{figure}

Figure \ref{self-white} shows the numerically computed self-correlation.  If we subtract the two-point correlation in Fig. \ref{white2compare} from that presented in this figure, we get the expected Gaussian. This is shown in Fig. \ref{white-sub} where it is compared with the analytical Gaussian function in Eq. (\ref{whitenn}) for 1 million events.

\begin{figure}[H]
\centering
\includegraphics[scale=0.5]{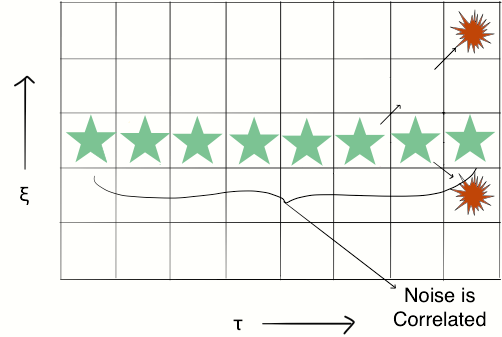}
\caption{Schematic of the self-correlation. The star denotes a noise source and the bursts are the charge fluctuations resulting from noise. (color online)}
\label{manystars}
\end{figure}

Now we move on to the meaning of self-correlation for colored noise. Based on the prescription of self-correlation that we discussed in the introduction, we consider the schematic diagram in Fig. \ref{manystars}.  We are interested in noise sources at one particular $\xi$ because noise generated at any other $\xi$ would be uncorrelated. 

Let us try to understand what the analytical formula for this would look like. We start with the following expression for the charge fluctuation in $k$-space.
\be 
\delta \tilde{n}(k,\tau) = -\frac{1}{\tau} \int_{\tau_i}^{\tau} d\tau' s(\tau') \tilde{G}(k,\tau,\tau') \tilde{f}(k,\tau') \,.
\ee
Here $\tilde{G}$ is the Green's function for the homogeneous part of the SDE (\ref{color-kspace}), which can be written down in terms of Kummer's function for the temperature dependences listed after that equation \cite{plumbergkapusta}.  This gives the full form of the two-point correlation function as in Eq. (49) of Ref. \cite{plumbergkapusta}.
\ba
\langle \delta n (\xi_1,\tau_f)\delta n(\xi_2,\tau_f)\rangle 
&=& \frac{1}{\tau_f^2}\int \frac{dk}{2\pi} e^{i k (\xi_1 - \xi_2)} \int d\tau's(\tau') \int d\tau''s(\tau'') \nonumber \\
&\times& \tilde{G}(k,\tau_f,\tau')\; \tilde{G}(-k,\tau_f,\tau'') \, {\cal N}(\tau',\tau'') \,.
\ea
Following Eqs. (54) and (55) of Ref. \cite{plumbergkapusta}, we can write the self-correlation term as
\ba
 \langle \delta n (\xi_1,\tau_f) &\delta  n&(\xi_2,\tau_f)\rangle_{\text{self}} \nonumber \\
 &=&  \frac{\chi_Q(\tau_f)T_f}{A \tau_Q} \int \frac{d\tau''}{\tau''} \left[ e^{-(\tau_f - \tau_2)/\tau_Q} - e^{-(\tau_f + \tau_2 - 2\tau_i)/\tau_Q} \right] \int \frac{dk}{2\pi} e^{i k (\xi_1 - \xi_2)} \frac{\tilde{G}(-k,\tau_f,\tau'')}{ik} \nonumber \\
 &=& \frac{s(\tau_f)}{D_Q} \int \frac{dk}{2\pi} e^{i k (\xi_1 - \xi_2)} 
\int d\tau'' s(\tau'') \; \frac{\tilde{G}(k,\tau_f,\tau_f)}{ik} \; \frac{\tilde{G}(-k,\tau_f,\tau'')}{ik} \, {\cal N}(\tau_f,\tau'') \,. \label{self_correlations}
\ea
Recall from Ref. \cite{plumbergkapusta} that $\tilde{G}(k,\tau_f,\tau_f) = ik$.  It denotes a noise fluctuation that was generated at the final time and didn't have to move anywhere.  The $ \tilde{G}(-k,\tau_f,\tau'')$ is a noise fluctuation generated at a time $ \tau'' < \tau_f$ and then moved to $\xi_2 $ at $\tau_f$. For white noise, fluctuations generated at two separate spacetime points can't be correlated, so the fluctuation generated at $\tau_f$ is only correlated to itself.  Hence for white noise, $\tau_Q \to 0$, and the self-correlation is just a Dirac $\delta$-function. As $\tau_Q$ increases, the more backward in time the noise sources would be correlated.  Once generated the noise will travel and give rise to a correlated electric charge fluctuation further away in spacetime rapidity. Hence we expect the self-correlation term to be more spread out in spacetime rapidity. One can use the same SDE solver for generating the self-correlation.  The only change is that the Green's function $\tilde{G}$ is replaced by $\tilde{G}/(ik)$ when solving for the charge density fluctuation.  

\ba
\langle &\delta n&(\xi_2,\tau_f) f(\xi_1,\tau_f)\rangle \nonumber \\
&=& \left\langle \int \frac{dk}{2\pi} e^{ik\xi_2} \delta \tilde{n}(k,\tau_f) \int \frac{dk_1}{2\pi}e^{ik_1\xi_1}\tilde{f}(k,\tau_f) \right\rangle \nonumber\\
&=& \frac{1}{\tau_f} \int^{\tau_f}_{\tau_i} d\tau' s(\tau') \int \frac{dk}{2\pi} e^{ik\xi_2} \frac{\tilde{G}(k,\tau',\tau_f)}{ik} \int \frac{dk_1}{2\pi} e^{ik_1\xi_1} \langle \tilde{f}(k,\tau') \tilde{f}(k_1,\tau_f) \rangle \nonumber \\
&=& \frac{1}{\tau_f}  \int^{\tau_f}_{\tau_i} d\tau' s(\tau') \int \frac{dk}{2\pi} e^{ik(\xi_2-\xi_1)} \frac{\tilde{G}(k,\tau',\tau_f)}{ik} \mathcal{N}(\tau',\tau_f) \nonumber \\
&=& \frac{D_Q}{s_f \tau_f} \, \frac{\chi_f T_f}{A\tau_Q}   \int^{\tau_f}_{\tau_i} \frac{d\tau'}{\tau'}  \left[ e^{-|\tau_f-\tau'|/\tau_Q} - e^{(2\tau_i-\tau_f-\tau')/\tau_Q} \right]  \int \frac{dk}{2\pi} e^{ik(\xi_2-\xi_1)} \frac{\tilde{G}(k,\tau',\tau_f)}{ik}
\ea

Note that in going to the second step from the first step, we have used $\tilde{G}/(ik)$ and not $\tilde{G}$. The implication is that for generating the self-correlations, we will be using the differential equation whose Green's function is going to be $\tilde{G}/(ik)$, instead of $\tilde{G}$. Thus, we arrive at the following relation for self-correlations.

\be
 \langle \delta n (\xi_1,\tau_f) \delta n(\xi_2,\tau_f)\rangle_{\text{self}} =
\frac{s(\tau_f) \tau_f}{D_Q} \langle \delta n (\xi_1,\tau_f) f(\xi_2,\tau_f)\rangle \, .
\ee
If $\tilde{G}/(ik)$ is our desired Green's function, then $Z \equiv (\tau \delta n(\xi,\tau))/(\tau_f s(\tau_f)) = \delta n(\xi,\tau)/s(\tau)$ satisfies the following equation
\be
\left( z^2  + 2 z \frac{\tau_Q}{\tau_i} \right) \frac{\partial Z}{\partial z} + z^2 \frac{\tau_Q}{\tau_i} \frac{\partial^2 Z}{\partial z^2} - v_Q^2  \frac{\tau_Q}{\tau_i} \frac{\partial^2 Z}{\partial \xi^2} 
+ \left( z + \frac{\tau_Q}{\tau_i} \right) f + z\frac{\tau_Q}{\tau_i} \frac{\partial f}{\partial z} = 0 \,,
\ee
where $z = \tau/\tau_i$.  This is the same as Eq. (\ref{diffeqcolor}) except that $\partial f/\partial \xi$ is replaced by $f$ and $\partial^2 f/\partial \xi \partial \tau$ is replaced by $\partial f/\partial \tau$.  The justification is discussed in the Appendix C.  

\begin{figure}[H]
\centering
\includegraphics[scale=0.13]{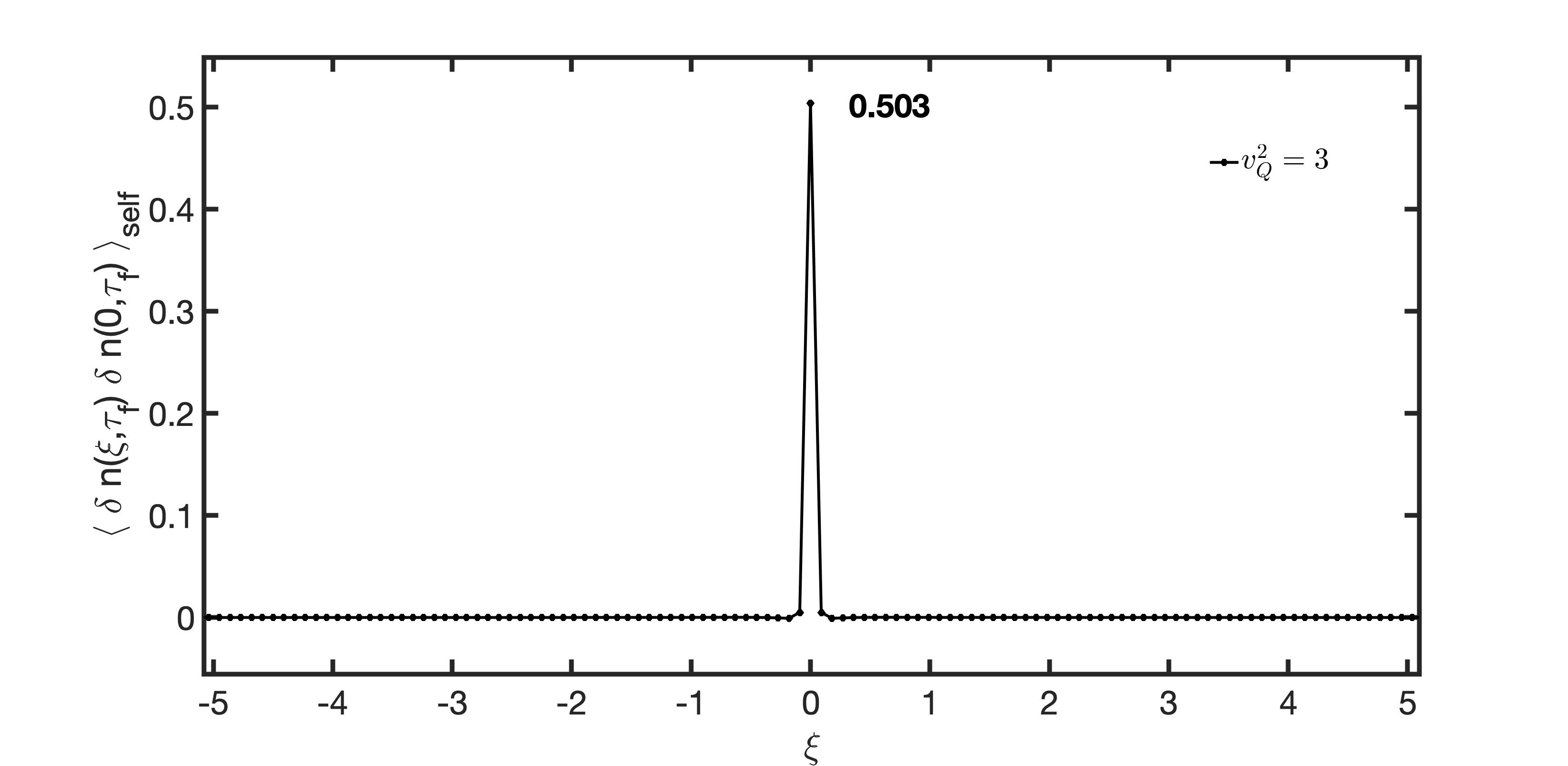}
\caption{Numerical results for self-correlation for white noise.}
\label{SelfFiguet3}
\end{figure}

In Fig. \ref{SelfFiguet2} we show the self-correlation at the final time $\tau_f$ for various values of $\tau_Q$. As the speed of propagation decreases, the height of the self-correlation decreases and widens.  As a check, the limit $\tau_Q \to 0$ is shown in Fig. \ref{SelfFiguet3}.
\begin{figure}[H]
\centering
\includegraphics[scale=0.13]{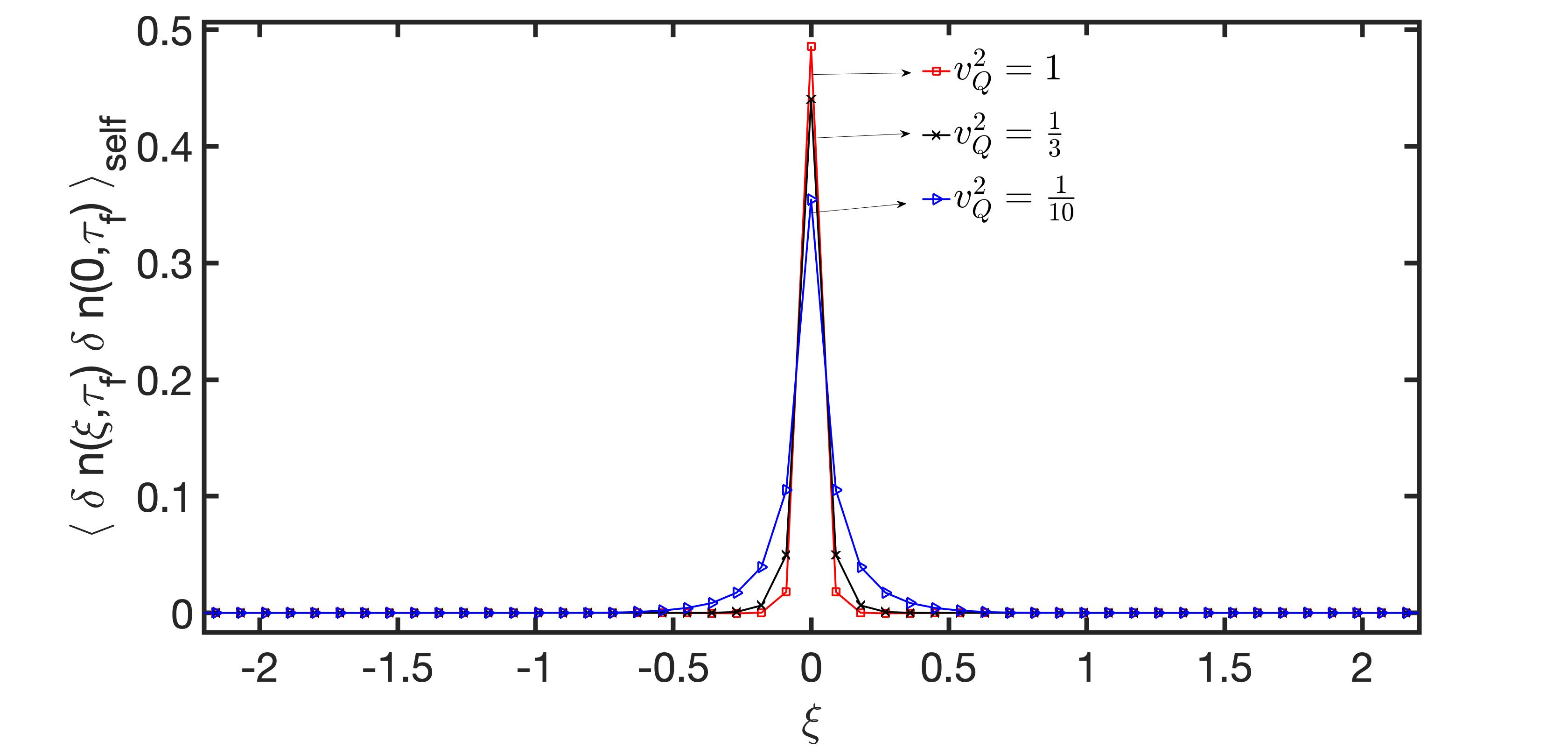}
\caption{Numerical results for self-correlations for colored noise.  (color online)}
\label{SelfFiguet2}
\end{figure}

\section{Balance Functions}
\label{balance}

Balance functions are described in Ref. \cite{spratt}. The width of a balance function plotted against particle rapidity is a measure of the diffusion. Balance functions have been experimentally studied by the ALICE and STAR collaborations \cite{balance1,balance2,balance3,balance4}. Reference \cite{ling_springer_stephanov} studied the effect of white noise in balance functions and compared their analytical results with experimental data. Reference \cite{plumbergkapusta} calculated balance functions for colored noise. We will see how the widths of balance functions change if we vary the speed of propagation of signals in case of Catteneo noise.

To see the effect of charge fluctuations in particle spectra we have to calculate how the fluctuations freeze-out. The freeze-out happens when the system has expanded and cooled to the extent that thermal equilibrium can no longer be maintained. Then hadrons freeze-out and free-stream to the detectors. The standard procedure to calculate freeze-out abundances of particles is the Cooper-Frye prescription \cite{cooper_frye}. This formula gives us the distribution of emitted particles on a freeze-out hypersurface $\Sigma_f$. This procedure has already been performed for this hydrodynamical model in Refs. \cite{kapusta2012,plumbergkapusta,ling_springer_stephanov,torres} We will just give the salient features of that calculation here.
\be
E\frac{dN}{d^3p} = d\int_{\Sigma_f} \frac{d^3\sigma_{\mu}}{(2\pi)^3} p^{\mu}f({\vec x, \vec p}) 
\ee
Here $d$ is the degeneracy of the particle species under consideration. We take the distribution function to be the relativistic Boltzmann
\be
f({\vec x, \vec p}) = e^{-(u \cdot p-\mu)/T} \,,
\ee
where $\mu$ is the chemical potential for that particle.  The energy flux through an infinitesimal freeze-out fluid cell is given by 
\be
 d^3 \sigma_{\mu} p^{\mu} = \tau_f \,  d\xi \, d^2 x_{\perp} m_{\perp}  \cosh(y - \xi) \,.
\ee
The variable $y$ represents the particle rapidity
\be 
p^{\mu} = (m_{\perp}\cosh y , p_x , p_y, m_{\perp}\sinh y )  
\ee
with $m_{\perp} = \sqrt{m^2+p_{\perp}^2}$ the transverse mass. 

The average number of particles per unit rapidity at the final freeze-out time is
\be
\left\langle \frac{dN}{dy} \right\rangle =  \frac{d A\tau_f T_f^3}{4\pi^2} \int^{\infty}_{-\infty} \frac{dx}{\cosh^2 x} 
\Gamma\left(3,\frac{m}{T_f}\cosh x \right) \,.
\ee 
Reference \cite{plumbergkapusta} calculates the fluctuation in this quantity due to a $\mu$ around the freeze-out $\mu_f = 0$. After a few more steps of algebra, the fluctuation in $\frac{dN}{dy}$ reads 
\be
\delta\left( \frac{dN}{dy} \right) =  \frac{d A\tau_f T_f^2}{4\pi^2} \int d\xi \, \delta n \, F_n(y-\xi)  
\ee
where $F_n$ is the smearing function
\be
 F_n(x) = \frac{1}{\chi_Q \cosh^2 x} \Gamma\left(3,\frac{m}{T_f}\cosh x \right) \,.
\ee
Using this in the definition of the charge balance function we arrive at the Eq. (74) in Ref. \cite{plumbergkapusta}.
\be
B(\Delta y) = \frac{\langle \delta \left( dN/dy_1 \right)  \delta \left( dN/dy_2 \right) \rangle }{\langle dN/dy \rangle} 
 = \frac{dA\tau_f T_f}{4\pi^2}  \frac{C(\Delta y)}{Q(m/T_f)} \,.
\ee
Here 
\be
C(\Delta y) = 2 \pi \int d\xi_1 d\xi_2 \, F_n(y_1-\xi_1) \, F_n(y_2-\xi_2) \, C_{nn}(\xi_1-\xi_2 , \tau_f) \,.
\ee
The two-point correlator for the charge fluctuation is $C_{nn}(\xi_1-\xi_2 , \tau_f)$ which is obtained from the solution of the SDE. The function $Q$ is given by 
\be
Q\left(\frac{m}{T_f}\right) = \int^{\infty}_{-\infty} \frac{dx}{\cosh^2 x} \Gamma\left(3,\frac{m}{T_f}\cosh x \right) \,.
\ee

\begin{figure}[H]
\centering
\includegraphics[scale=0.13]{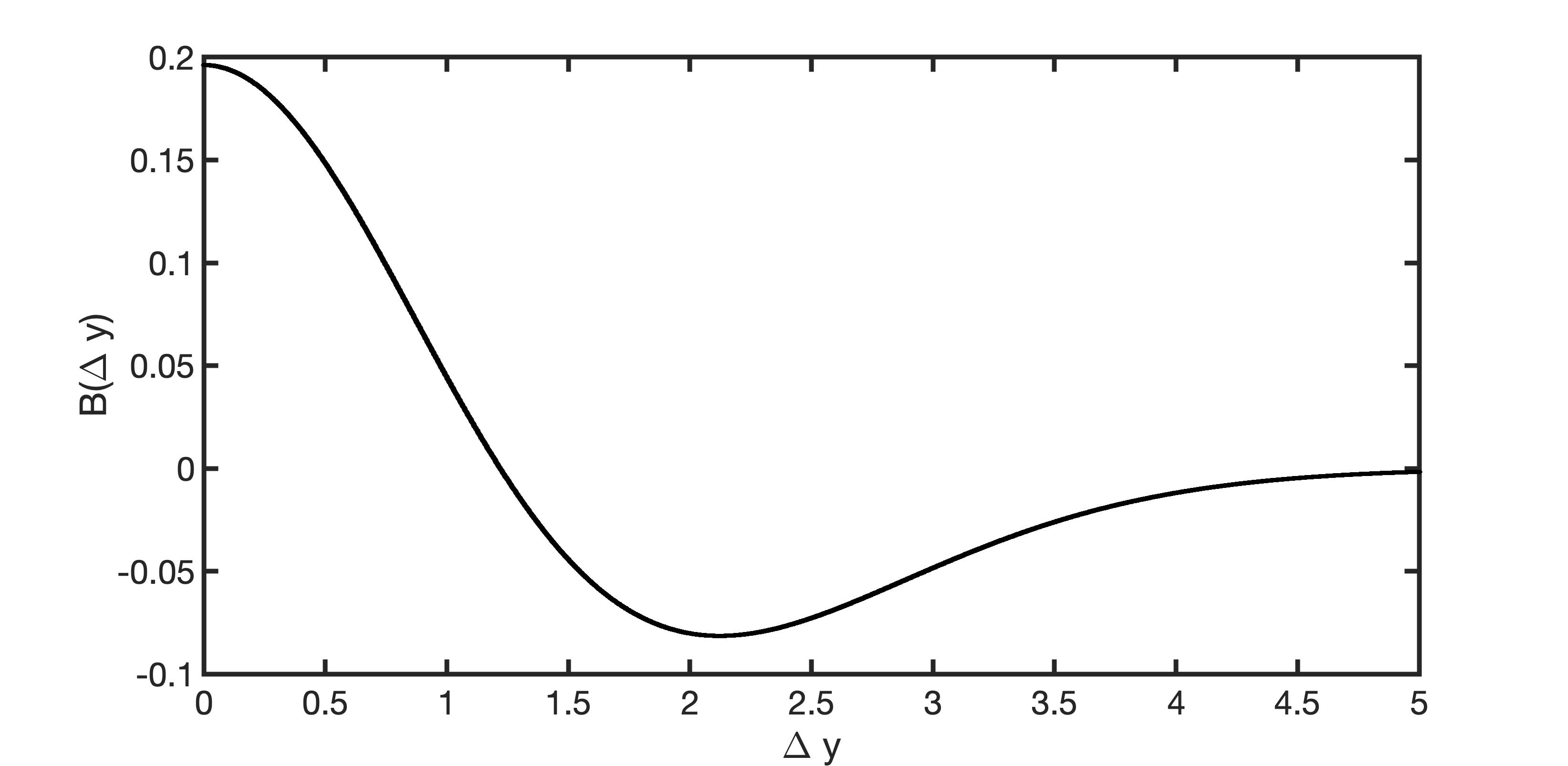}
\caption{Balance function for the full white noise two-point function.}
\label{Bwhitefull}
\end{figure}

\begin{figure}[H]
\centering
\includegraphics[scale=0.13]{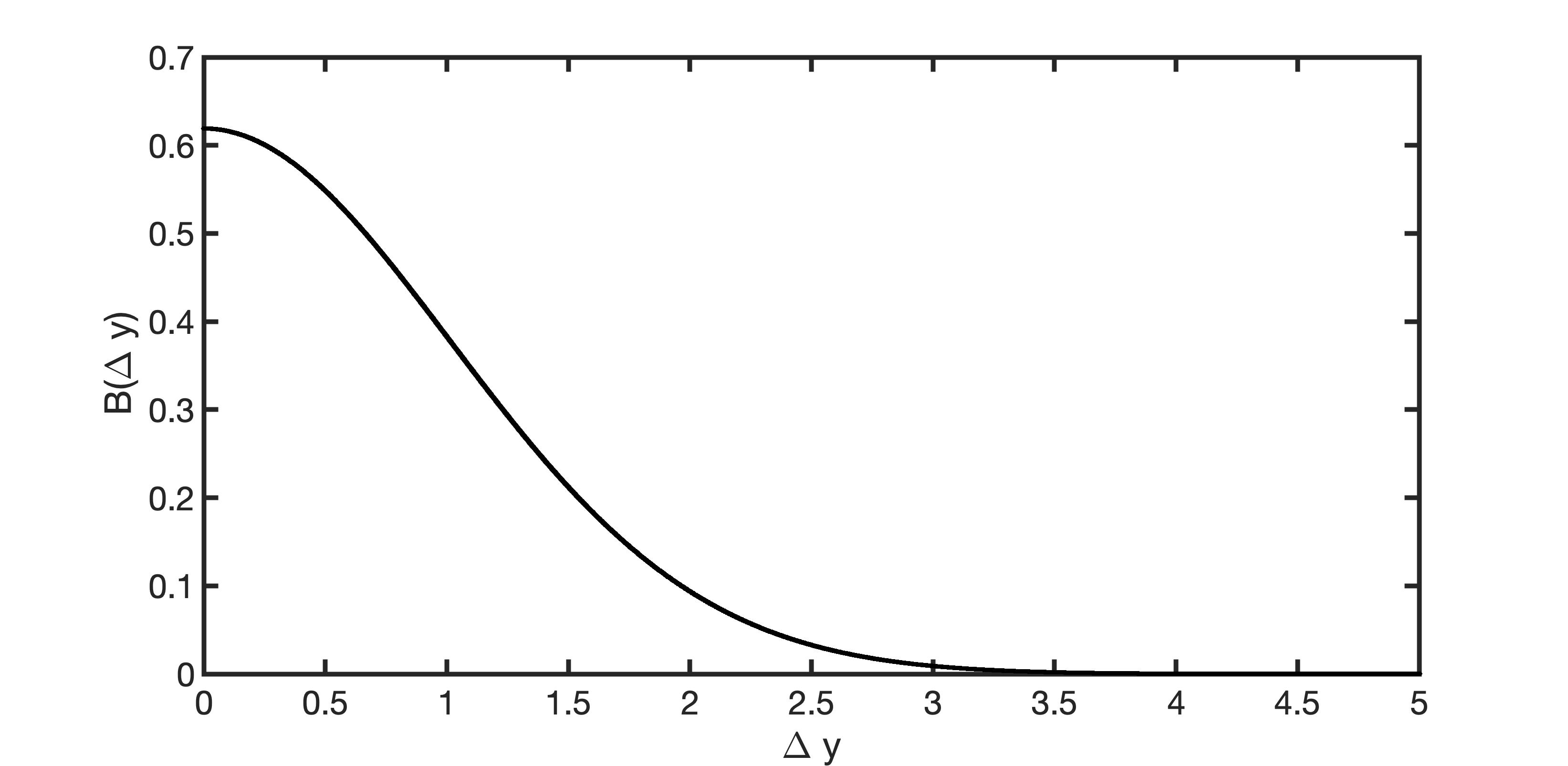}
\caption{Balance function for the self-correlation of white noise.}
\label{Bwhiteself}
\end{figure}

\begin{figure}[H]
\centering
\includegraphics[scale=0.13]{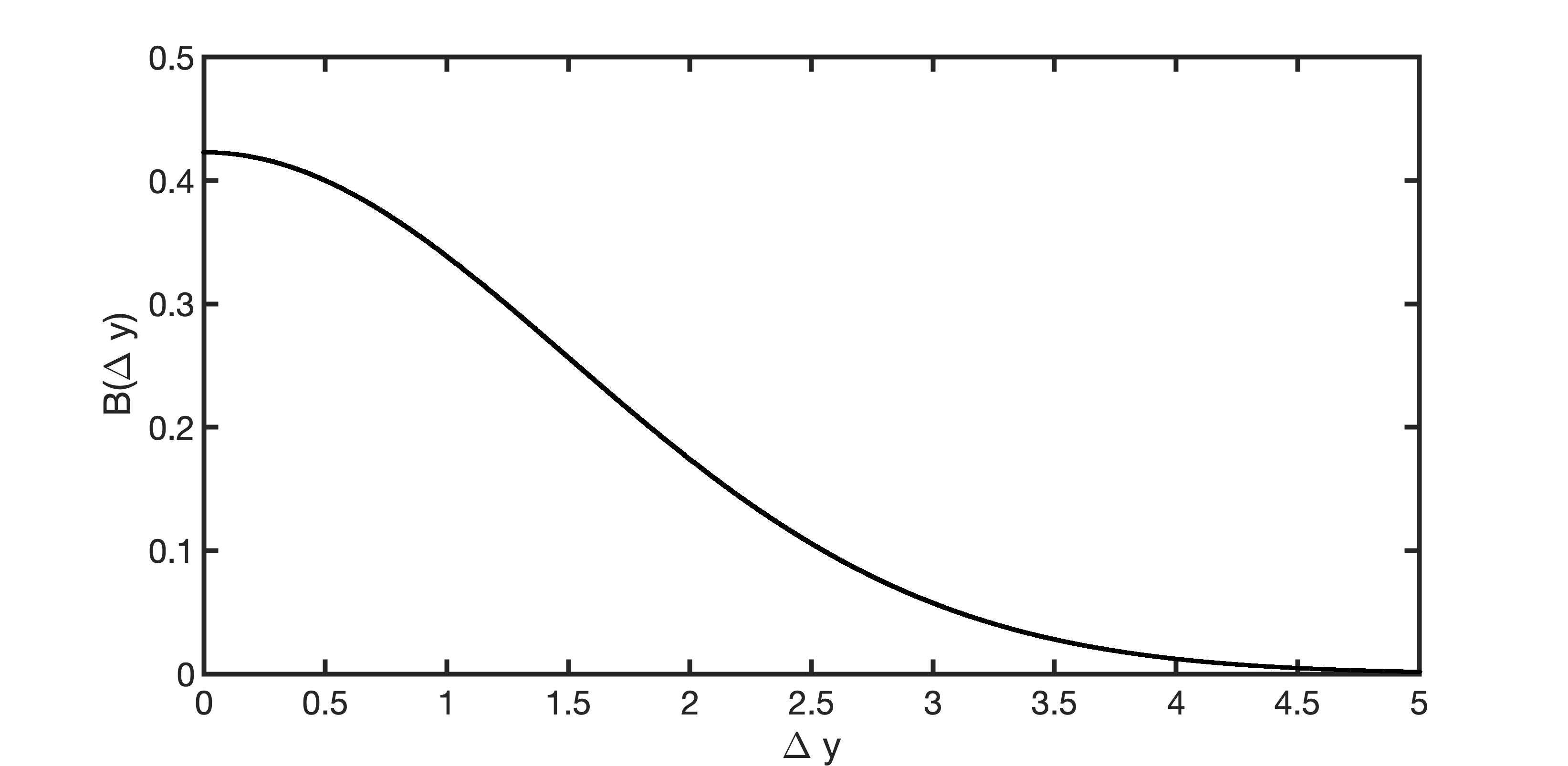}
\caption{Balance function for the pure two-point function of white noise.}
\label{Bwhitesub}
\end{figure}

Let us first demonstrate the trivial self-correlation for white noise in terms of the balance function for pions.  Figure \ref{Bwhitefull} shows the balance function for the full unsubtracted correlation function for white noise.  Notice the positive and negative part of the curve; this is because the full two-point correlation for white noise is composed of a positive self-correlation and a negative piece which does not include any self-correlation. The balance function for the self-correlation part only is shown in Fig. \ref{Bwhiteself}. When this is subtracted from Fig. \ref{Bwhitefull} one obtains the so-called subtracted balance function shown in Fig. \ref{Bwhitesub}.

\begin{figure}[H]
\centering
\includegraphics[scale=0.13]{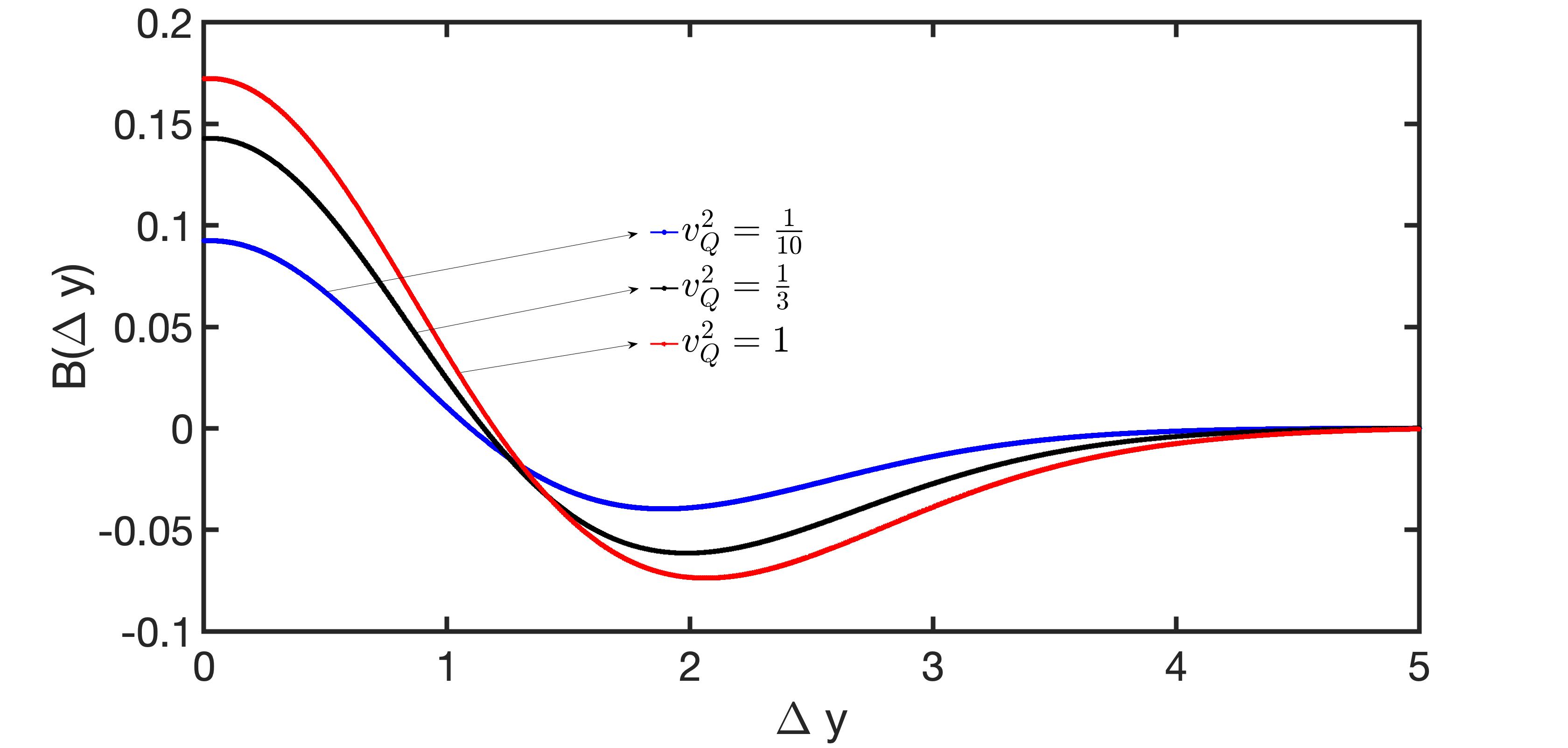}
\caption{Balance function for the full unsubtracted two-point function. (color online)}
\label{Bcolorfull}
\end{figure}

\begin{figure}[H]
\centering
\includegraphics[scale=0.13]{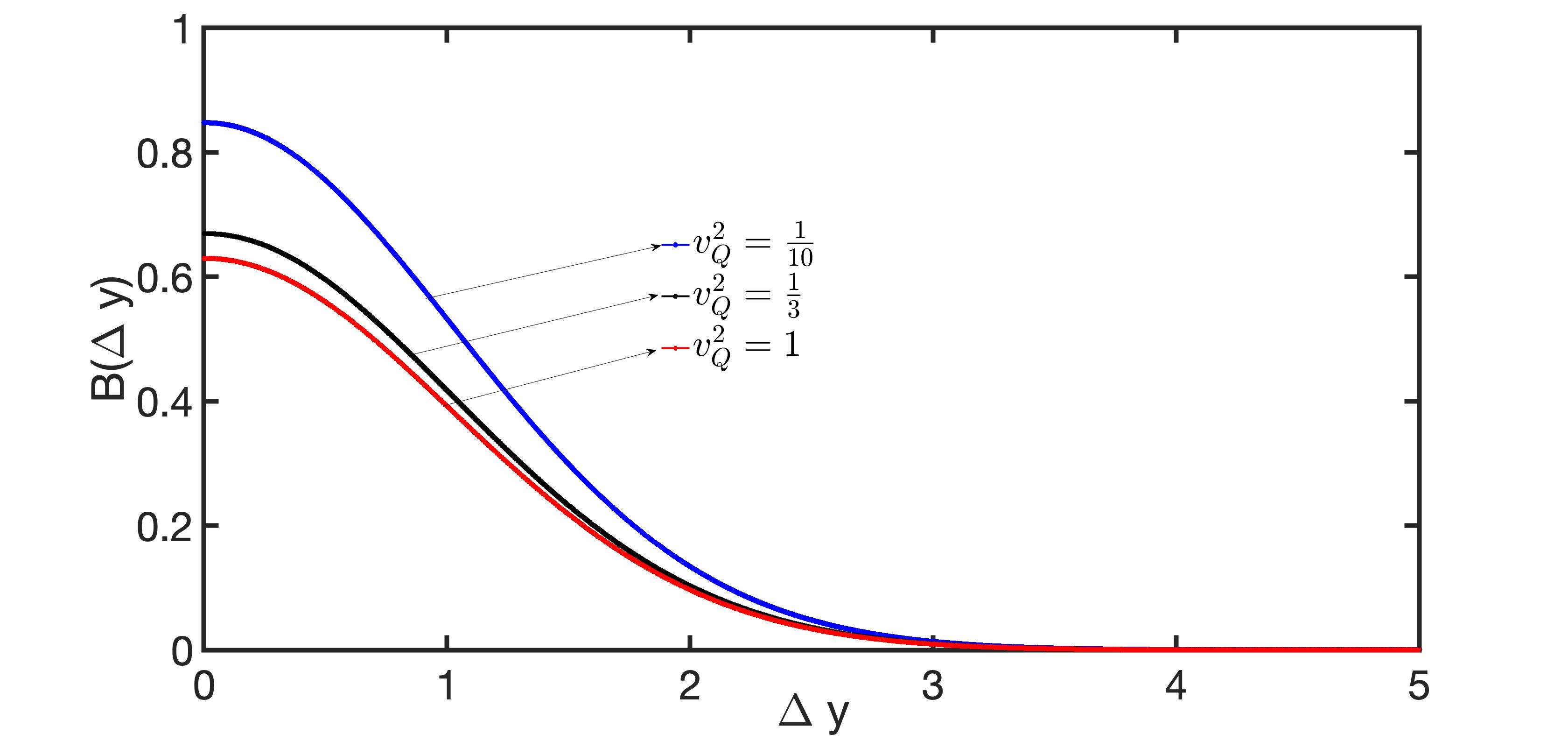}
\caption{Balance function for the self-correlation part of two-point function. (color online)} 
\label{Bcolorself}
\end{figure}

\begin{figure}[H]
\centering
\includegraphics[scale=0.13]{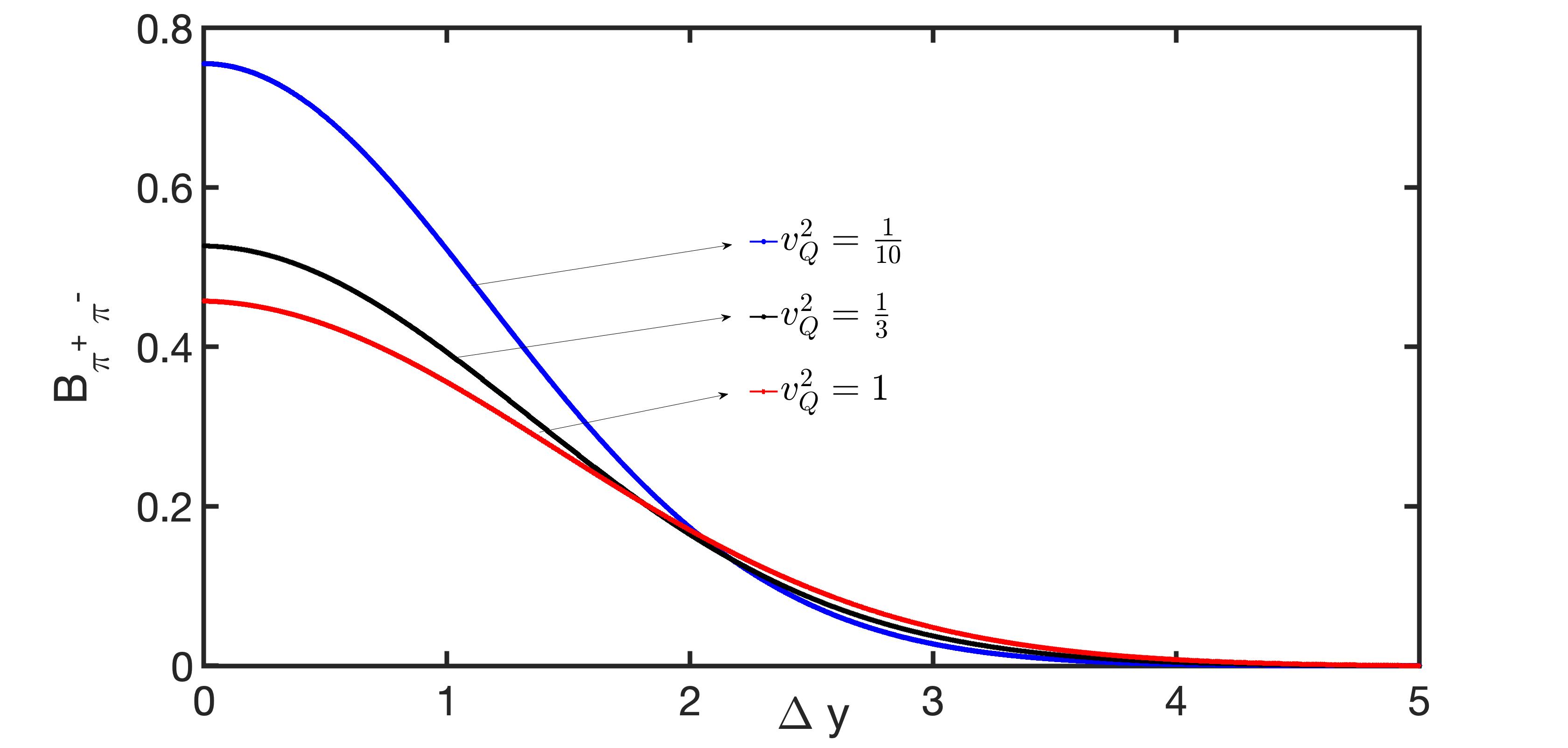}
\caption{Balance function for the subtracted two-point correlation function. (color online)}
\label{Bcolorsub}
\end{figure}

We follow the same procedure for carrying out cancellations of the contributions arising from the self-correlations for colored noise to the balance function. Figure \ref{Bcolorfull} shows the full unsubtracted balance functions for various values of $v_Q$ for colored noise.  Figure \ref{Bcolorself} shows the self-correlation part only, and Fig. \ref{Bcolorsub} shows the subtracted balance functions.  The width of the subtracted balance function denotes the diffusion distance.  That width increases with increasing $v_Q$, as expected, since it represents the rapidity interval over which the average charge pair has diffused to by freeze out.

We estimated the error in our numerical simulations using the jackknife method. This method estimates the error of statistics without making any assumptions about the distribution that generated the data. It only uses the sample provided. We create jackknife samples over the whole data set which are ``leave-one-out" data sets. In our case, we consider the two-point correlation statistic $S$ on the original sample size of $10^7$ events. We leave out the $i_{th}$ event to create the $i_{th}$ jackknife statistic $S_i$. The average of the jackknife sample is $S_{\rm avg} = \sum_i S_i/n$. The jackknife error is then estimated as 
\be
\sigma_{\rm jack} = \sqrt{\frac{n-1}{n} \sum_i (S_i-S_{\rm avg})^2}
\ee

\begin{figure}[H]
\centering
\includegraphics[scale=0.13]{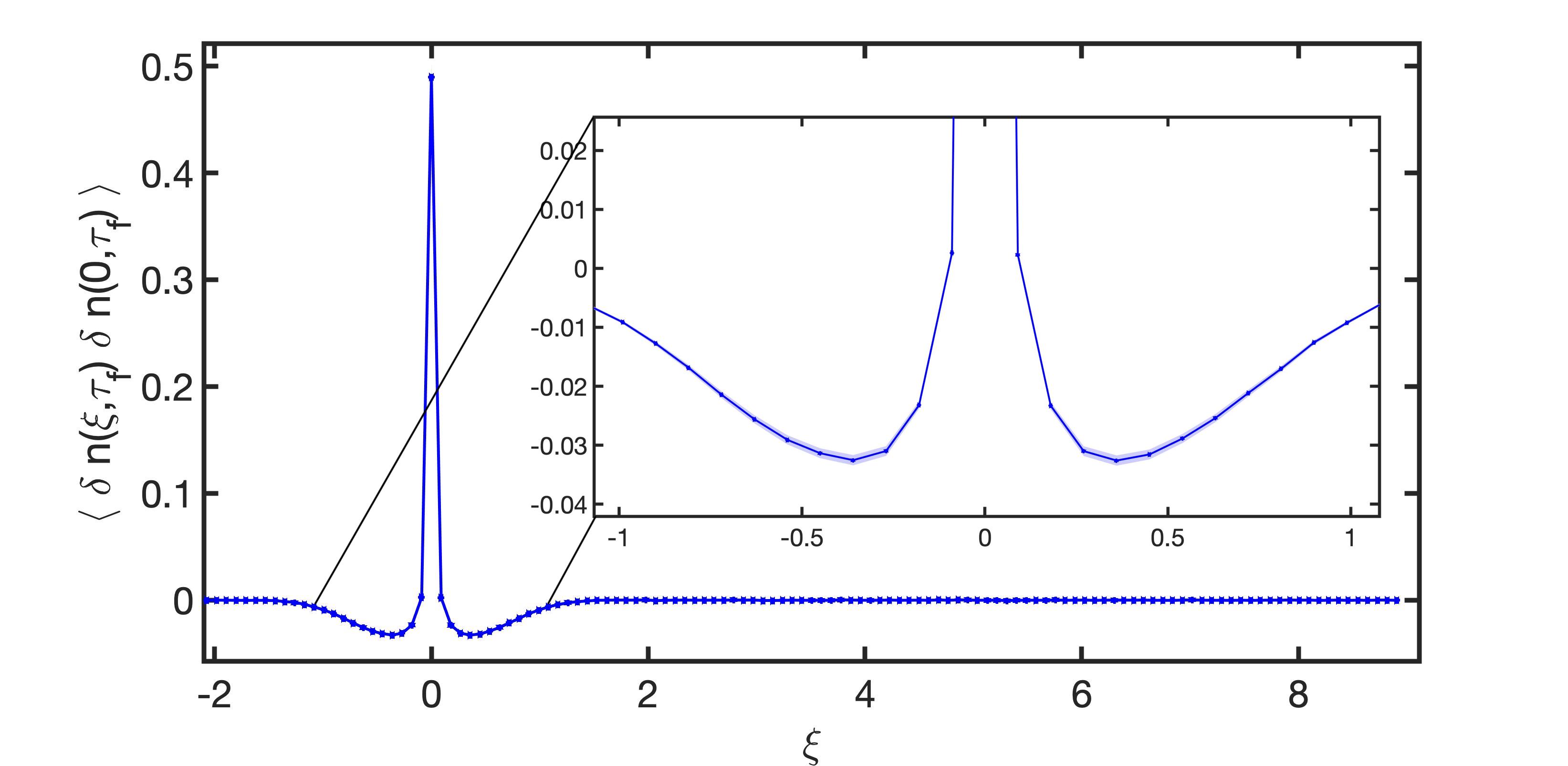}
\caption{Jackknife error bounds for $v_Q^2 = 1/10$. (color online)}
\label{jackknife}
\end{figure}

The error we observe on $\langle \delta n \delta n \rangle $ is of the order of $10^{-2}\; \text{MeV}^3 \: \text{fm}^{-3}$. This amounts to $\sigma_{\langle \delta n \delta n \rangle} / \langle \delta n \delta n \rangle \approx 10^{-3}$.
We give a representative plot of the error bounds for $v_Q^2 = 1/10$ in Fig. \ref{jackknife}. The bounds are visible only when zoomed in. This shows that for $10^7$ events, the statistical error in our simulations turn out to be negligible.
 
\section{Conclusions}
\label{conclude}

State-of-the-art modeling of high energy nuclear collisions uses relativistic 2nd order viscous hydrodynamics.  The fluctuation-dissipation theorem says that viscosity and thermal fluctuations are intricately connected.  Although thousands of particles are produced in these collisions, that is still immensely smaller than Avogadro's number.  Therefore it has become apparent that thermal fluctuations really ought to be part of the standard model for heavy ion collisions \cite{kapusta2012}.  Fully 3+1 dimensional hydrodynamic simulations are required which presents a major challenge for implementation of thermal noise.  The goal of this paper is to understand the numerical methods necessary to do this and, furthermore, how to subract self-correlations from the numerically computed two-point correlators in order to compare with experiment.  We chose to study causal electric charge diffusion in the boost-invariant 1+1 dimensional Bjorken model for two reasons.  First, a Cattaneo description of diffusion propagates signal at a finite speed which is a necessity in heavy ion collisions.  Second, this simple model was studied and solved with essentially analytic methods \cite{plumbergkapusta} against which we can compare to verify the validity of the purely numerical approach.   

We introduced the noise term in the dissipative charge conservation equation, which in our case is the Catteneo noise.  We simulated the stochastic differential equations that arise from the electric charge conservation equations. The way we solve the stochastic differential equations is by using normal random number generators with a specific, well-defined variance and then interpreting the derivatives of the noise in terms of what they mean when integrating by parts. The whole machinery on how to handle the noise is discussed in Appendix A. We used this methodology in simulating the white noise charge conservation equation and obtained the expect result.  Then we generated colored Catteneo noise using a Langevin equation.  We solved the full colored noise charge conservation equation and again obtained the expected result. The two-point charge correlator consists of two pieces. The first is the self-correlation, which is a manifestation of the stochastic nature of the dynamics.  Once this piece is subtracted off, we are left with the physically relevant two-point correlation function. The self-correlation is a trivial Dirac $\delta$-function in the case of white noise, but is more complicated for colored noise. In this work, we gave a physically insightful interpretation of the meaning of self-correlation in the case of colored noise. This interpretation allows us to use the stochastic differential equation solver we developed to generate the self-correlations. 
 
In the case of the white noise, we populated the whole spacetime lattice with noise source terms uncorrelated to each other. It is obvious that all the individual noise terms are not required to calculate the final two-point correlation function, but more than a single noise term is necessary. Hence Monte-Carlo simulations will be insufficient to reproduce the results for colored noise. One can, however, speed up the stochastic differential equation solving procedure by removing noise terms that are outside the causal past of the spacetime points for which we want to calculate the twopoint correlations. 
 
We used the results obtained to compute the balance functions for pions within the context of this model. As one would expect, reducing the speed of propagation of signal leads to narrowing of the balance functions and to a corresponding increase in their height at small rapidities. As done previously in Ref. \cite{plumbergkapusta} we neglect the contributions from resonance decays to the measured particle spectra used in the balance functions.  Our results are in good quantitative agreement with that previous study.  The numerical method used in this paper is verified.
 	
Future work entails furthering the current methodology to a full 3+1 dimensional fluid dynamical models of heavy ion collisions such as MUSIC \cite{music}. Further, the prescription for self-correlations given in this paper for Catteneo noise can be straightforwardly extended to the case of shear and bulk viscosity and thermal conductivity, the details of which are deferred to a future work.  Since baryon charge conductivity diverges near a critical point, this study can be extended to study charge fluctuations near the purported QCD critical point, which is also deferred to future work.  Another possible direction of future work would be to study the higher order cumulants in the presence of colored noise. Since two-point correlations and higher order cumulants are expected to diverge near a QCD critical point, the ultimate culmination of the present work would be to characterize the noisy hydrodynamics of near-critical point behavior of heavy ion collisions.
 
\section{Acknowledgements}
A. D. thanks Gaurav Nirala for enlightening discussions. We thank Chun Shen for suggesting the jackknife method. This work was supported by the U.S. DOE Grant No. DE-FG02-87ER40328.  C. P. acknowledges support from the CLASH project (KAW 2017-0036).  The authors acknowledge the Minnesota Supercomputing Institute (MSI) at the University of Minnesota for providing resources that contributed to the research results reported within this paper.

\appendix
\section{Numerical Simulation of SDEs}
In this appendix, we show our procedure for representing the Dirac delta function and its derivatives on a discrete lattice.

White noise is defined as $\langle f(x)f(x') \rangle = \delta(x-x')$ and $\langle f(x) \rangle = 0$. This implies that
\be
 \int dx \: g(x) \langle f(x)f(x')\rangle  = g(x')
\ee
On a discrete lattice, the integral becomes a sum over lattice points and $dx$ becomes the lattice spacing $\Delta x$. Hence
\begin{equation}
g(x_i) =\sum_{i'} g(x_{i'}) \langle f(x_i)f(x_{i'})\rangle \Delta x = \sum_{i'} g(x_{i'}) \left( \frac{\delta_{ii'}}{\Delta x} \right)\Delta x 
\end{equation}
From the above, we can conclude
\be 
\langle f(x_i)f(x_{i'})\rangle = \frac{\delta_{ii'}}{\Delta x} 
\ee

The $\delta_{ii'}/\Delta x$ becomes a Dirac-delta function in the limit $\Delta x \rightarrow 0$ which is the continuous case. Therefore we sample the white noise function $f$ from a Normal distribution of mean $0$ and standard deviation $1/\sqrt{\Delta x} $.
We use a random number generator for a large number of instances ($10^6 $) and compute the two-point function. It gives us the variance as the peak of a Kronecker delta at $x=0$. This is illustrated in the following figure. We used $\Delta x = 0.09$.

\begin{figure}[H]
\centering
\includegraphics[scale=0.1]{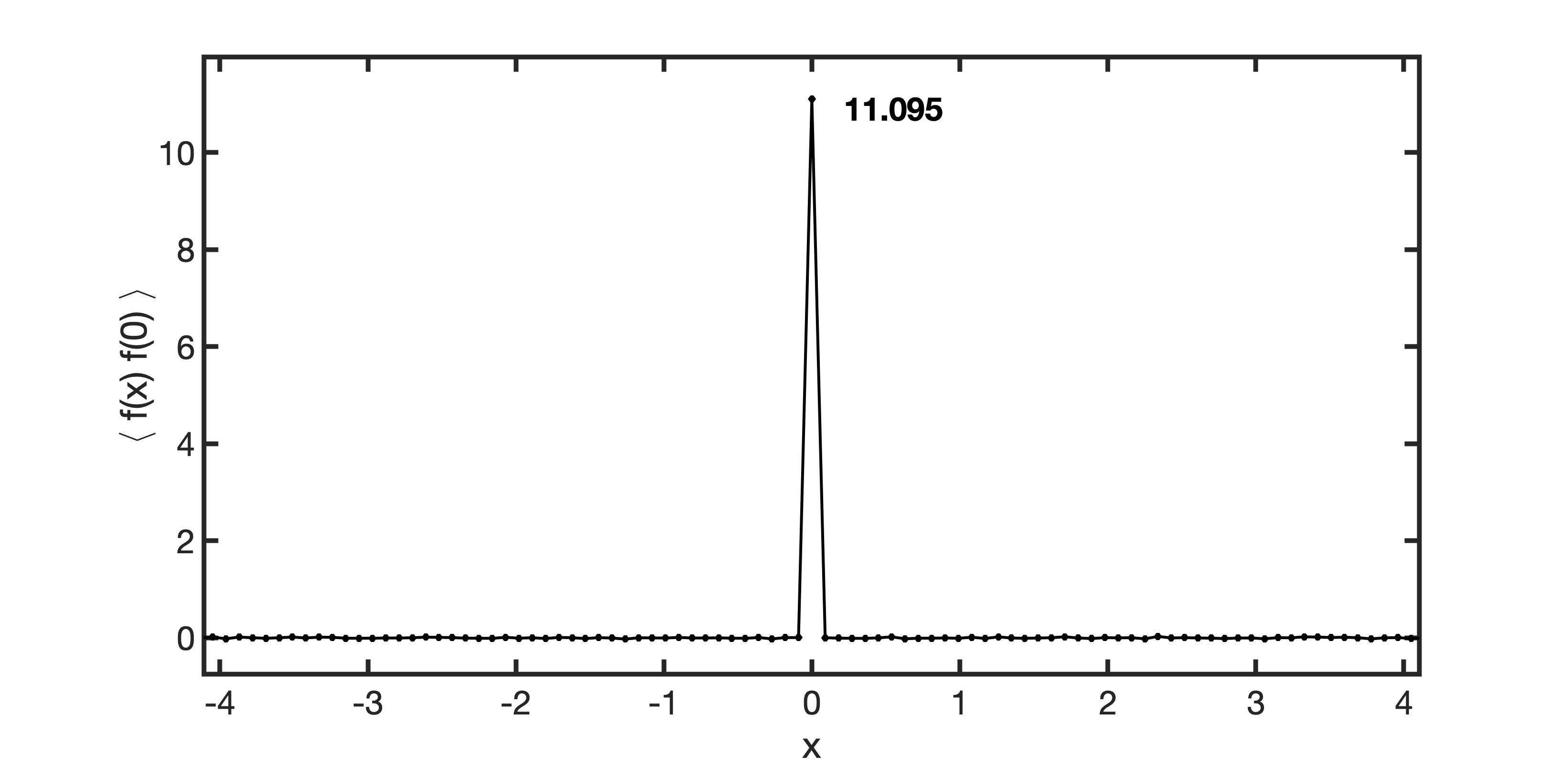}
\caption{Two-point function of $f$ for 1 million events.}
\end{figure}

Next, we investigate the correlation between white noise $f$ and its derivative $df/dx$. 
The two-point function $\langle f(x)f'(x')\rangle$ must then satisfy
 
\be 
  \int dx \, g(x)\langle f(x)f'(x')\rangle  = \int dx \, g(x) \frac{\partial }{\partial x'}  \delta(x-x')
= \frac{\partial }{\partial x'} \int dx\, g(x) \delta(x-x') = g'(x')
\ee
The derivative is $g'(x) = (g_{i+1}-g_{i})/\Delta x$ in a discrete lattice. Replacing the integral by the sum, we get
\be
g'(x_i) = \sum_{i'} g(x_{i'}) \langle f(x_i)f'(x_{i'})\rangle \Delta x =  \frac{g_{i+1}-g_{i}}{\Delta x} = \sum_{i'} g(x_{i'}) \left(\frac{\delta_{i+1,i'}-\delta_{i,i'}}{\Delta x^2} \right) \Delta x 
\ee
Hence
\be
 \langle f(x_i)f'(x_{i'})\rangle = \frac{\delta_{i+1,i'}-\delta_{i,i'}}{\Delta x^2} 
 \ee

\begin{figure}[H]
\centering
\includegraphics[scale=0.1]{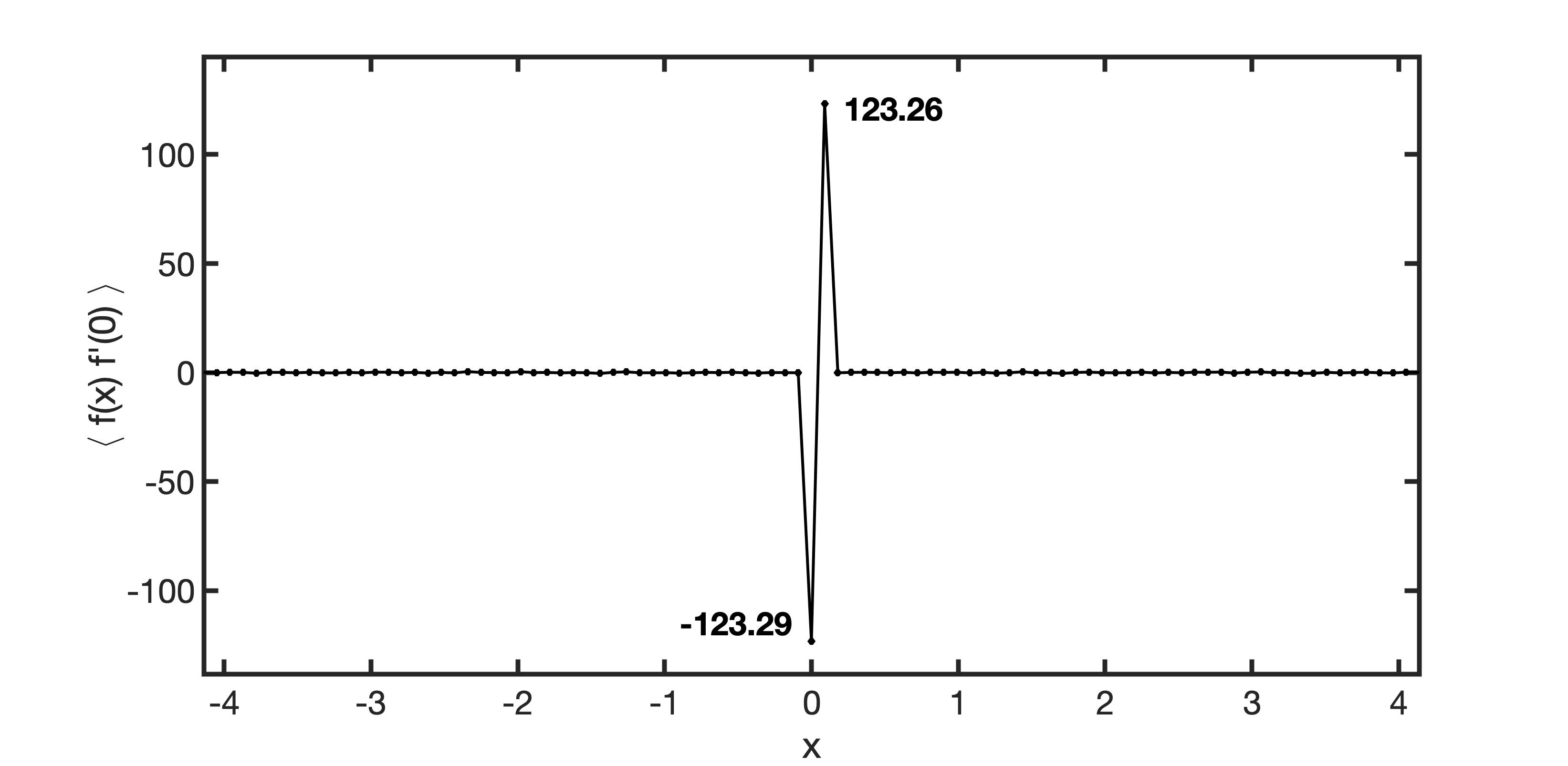}
\caption{The correlation between $f$ and its derivative for 1 million events.}
\label{whitederivative}
\end{figure}

If we again use the previous random number generator and calculate the two point function we get the results shown in Fig. \ref{whitederivative}.

\begin{figure}[H]
\centering
\includegraphics[scale=0.1]{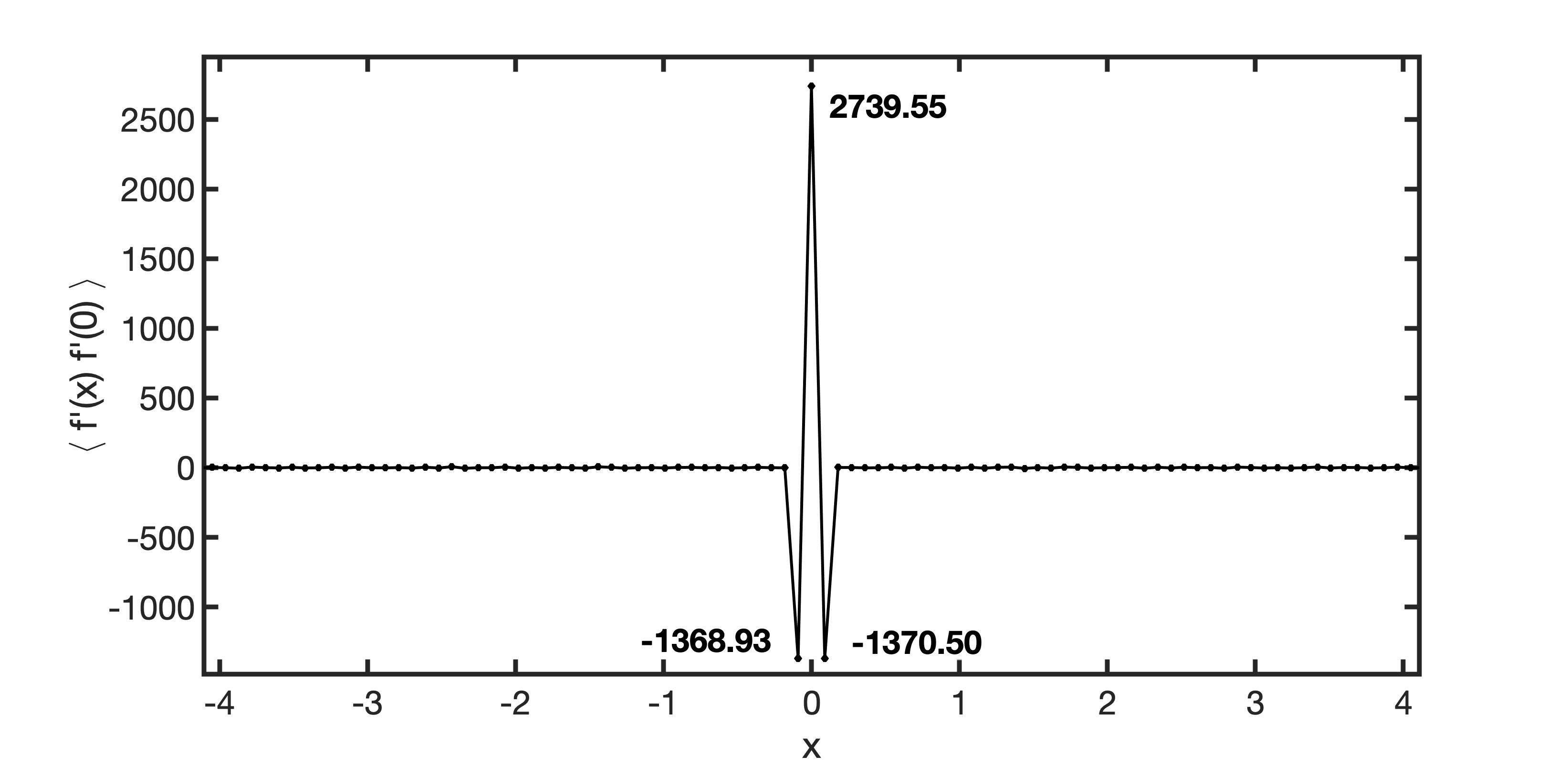}
\caption{The two-point correlation of the derivative of $f$ for 1 million events.}
\label{whitederivative2}
\end{figure}

Similarly, we can look into the correlation of the derivative of white noise with itself.
\be
\langle f'(x)f'(x')\rangle = \frac{\partial^2}{\partial x \partial x'} \delta(x-x')
\ee

\be
 \int dx \, g(x) \langle f'(x)f'(x')\rangle = \int dx \,g(x) \frac{\partial^2}{\partial x \partial x'} \delta(x-x') 
 = - \int dx \,g(x)\frac{\partial^2}{\partial x^2}  \delta(x-x') = -g''(x') 
\ee
In the second step we performed an integration by parts. The second derivative is defined in the discrete case as $g''(x) = (g_{i+1}+g_{i-1}-2g_{i})/\Delta x^2$. Substituting the discrete sum in place of the integral, we get
\be
-g''(x')  =  - \sum_{i} g_i \frac{\delta_{i,i'+1}+\delta_{i,i'-1}-2\delta_{i,i'}}{\Delta x^3} \Delta x = \sum_{i'} g(x_{i'}) \langle f'(x_i)f'(x_{i'})\rangle \Delta x
\ee
\be
 \langle f'(x_i)f'(x_{i'})\rangle = -\frac{\delta_{i,i'+1}+\delta_{i,i'-1}-2\delta_{i,i'}}{\Delta x^3}
 \label{dfdx_dfdx_2PF}
\ee
Figure \ref{whitederivative2} shows what we get numerically.

Integration of white noise is called a random walk $W(z)$ which is a succession of random steps as a function of $z$.  It is defined by 
\be
W = \int_{z_i}^z f(z) dz
\ee

We can easily calculate the variance of $W$:

\begin{equation}
\langle W^2(z) \rangle = \int_{z_0}^{z} dz' \int_{z_0}^{z} dz'' \langle  f(z')f(z'') \rangle = \int_{z_0}^{z} dz' \int_{z_0}^{z} dz''\delta(z'-z'')  =  \int_{z_0}^{z} dz' = z-z_0 
\end{equation}
On a discrete lattice, $W_{i+1} = W_i + f \Delta z$ where we source $f$ from a normal distribution of mean $0$ and standard deviation $1/\sqrt{\Delta z}$.
\be
 W_n = \sum_{i=1}^n f\Delta z
\ee 
 Hence the variance is
\begin{eqnarray}
 \langle W^2 \rangle
 	= \langle (\sum_{i=1}^n \Delta z f)^2 \rangle 
 	= \langle \sum_{i=1}^n (\Delta z f)^2 \rangle 
 	= \sum_{i=1}^n (\Delta z)^2 \langle f^2  \rangle
 	= \sum_{i=1}^n (\Delta z)^2 \frac{1}{\Delta z}
 	= \sum_{i=1}^n \Delta z
 	= z-z_0 \nonumber\\
\end{eqnarray}

\begin{figure}[H]
\centering
\includegraphics[scale=0.1]{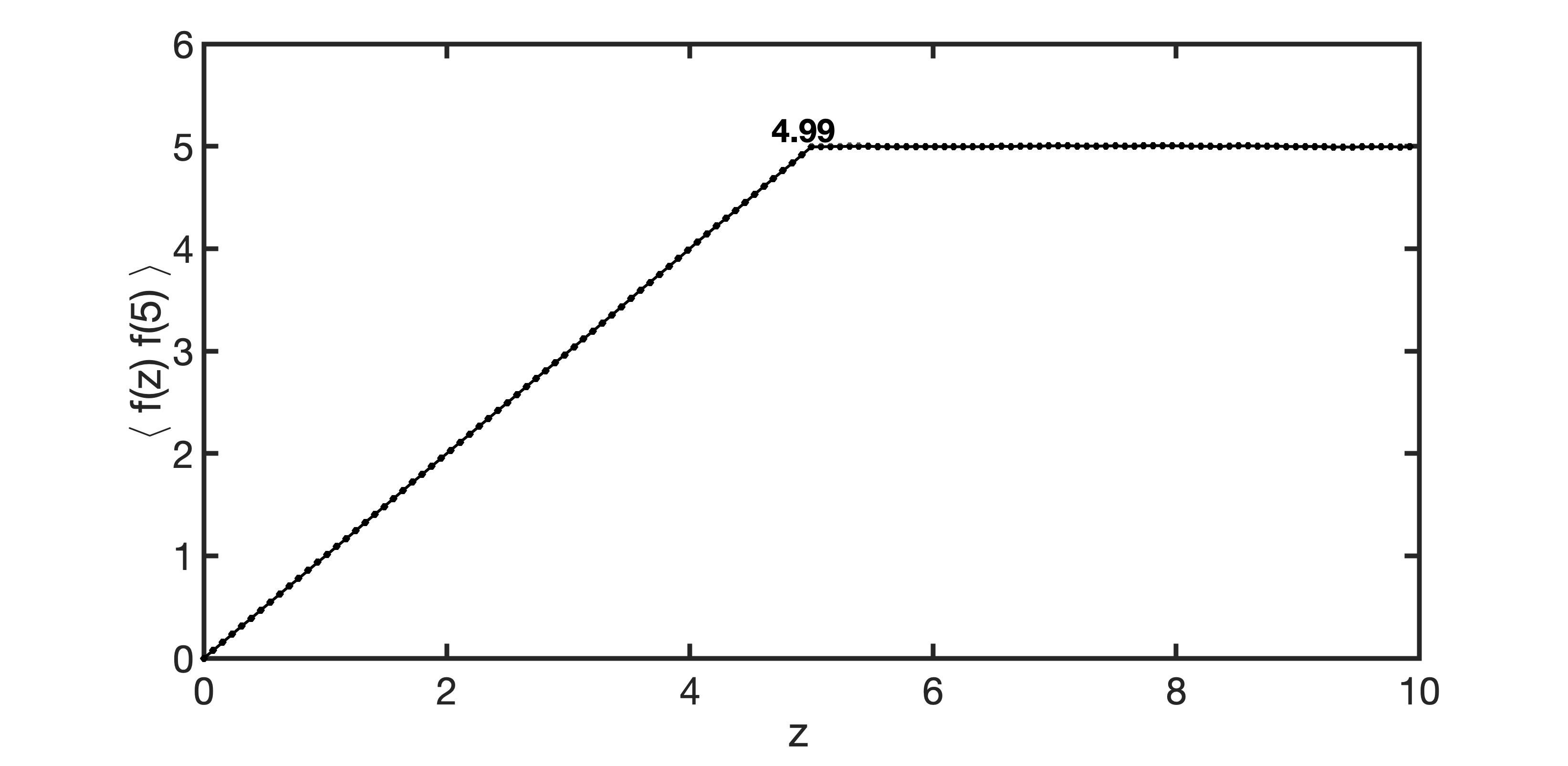}
\caption{Two point correlation of $W$ for 1 million events with $z - z_i = 5$.}
\label{rnwalk}
\end{figure}

Figure \ref{rnwalk} shows the numerical results.

\begin{figure}[H]
\centering
\includegraphics[scale=0.1]{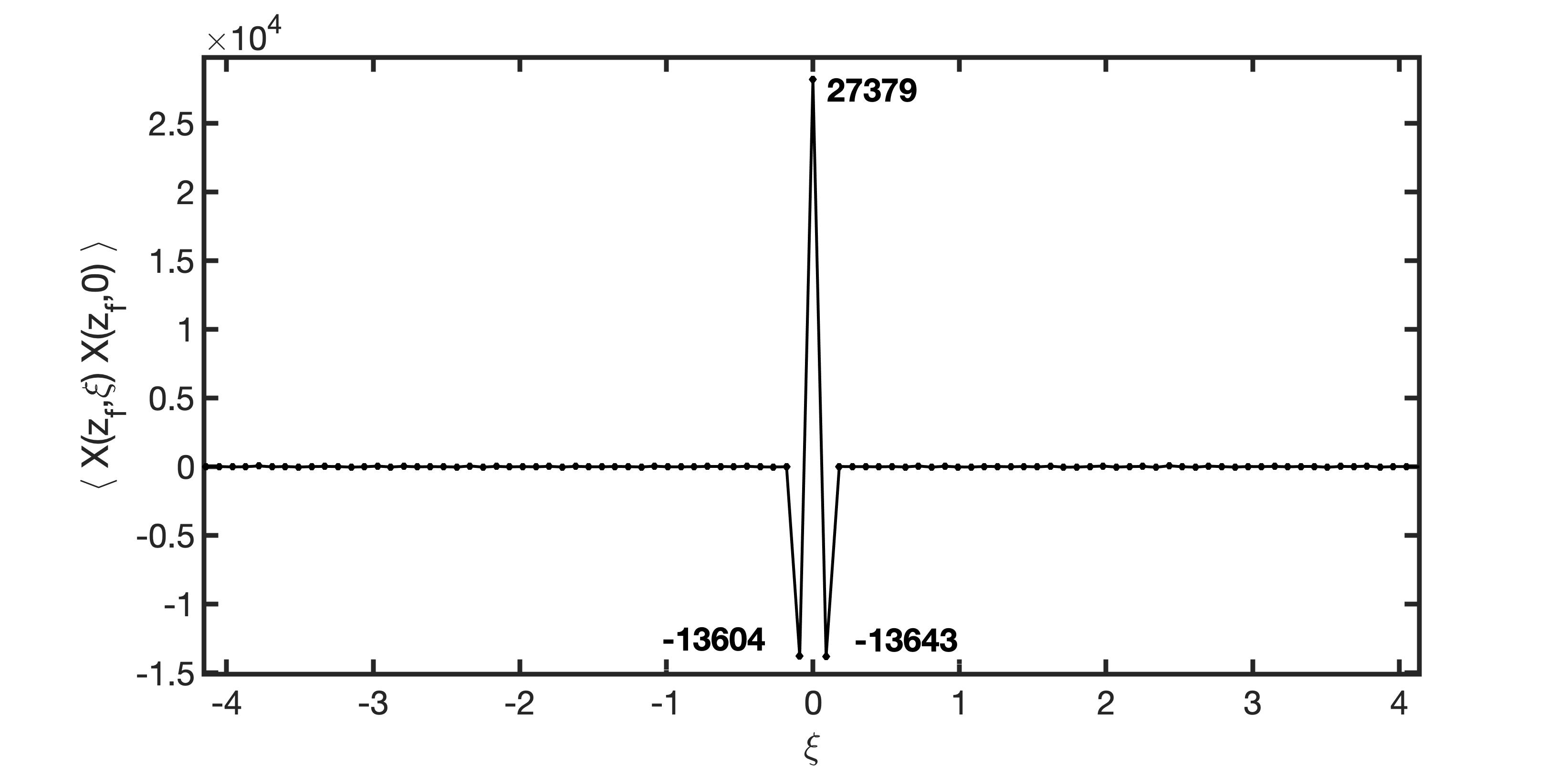}
\caption{Two point correlation of $X$ for a million events.}
\label{sde}
\end{figure}

We are ready to take up a simple stochastic differential equation to solve. Consider
\begin{equation}
\frac{dX}{dz} = -\frac{\partial f}{\partial \xi}
\end{equation}
where $z$ has dimensions of time and $\xi$ is dimensionless. Let us define the following two-point function 
$$ \langle f(z_1,\xi_1)f(z_2,\xi_2) \rangle  = M \delta(z_1-z_2)\delta(\xi_1-\xi_2) $$
Here $M$ has dimensions of time to make $f$ dimensionless. We calculate the two-point function in $\xi$.

 \begin{eqnarray} 
\langle X(z_f,\xi_1) X(z_f,\xi_2)\rangle  &=& \l\langle \int^{z_f}_{z_i} \frac{\partial f}{\partial \xi}(\xi_1) dz\int^{z_f}_{z_i} \frac{\partial f}{\partial \xi}(\xi_2) dz' \r\rangle \nonumber\\
&=& \int^{z_f}_{z_i}\int^{z_f}_{z_i} dz dz' \l\langle \frac{\partial f}{\partial \xi}(z,\xi_1) \frac{\partial f}{\partial \xi}(z',\xi_2) \r\rangle   \nonumber\\
&=& - \int^{z_f}_{z_i}\int^{z_f}_{z_i} dz dz' M \left( \frac{\delta_{i+1} + \delta_{i-1}- 2\delta_i}{\Delta \xi^3}\right) \delta(z-z')  \nonumber\\
&=& - M (z_f - z_i) \left( \frac{\delta_{i+1} + \delta_{i-1}- 2\delta_i}{\Delta \xi^3}\right)
\end{eqnarray}
We used Eq. \eqref{dfdx_dfdx_2PF} in the above calculation.
The two-point function has dimensions of time-squared and so is the expression on the right. On a discrete lattice,
\be
X(z+\Delta z) = X(z) - \Delta z \times \frac{\Delta f}{\Delta \xi} 
\ee
Figure \ref{sde} shows the numerical results using $z_f - z_i = 10$. 

\section{Analytical Solution of Langevin Equation}

The Langevin equation can be written as
\be
\frac{df(\tau)}{d\tau} = -\frac{1}{\tau_Q}f(\tau) + \frac{1}{\tau_Q}\zeta(\tau)
\ee
Here $\zeta$ is white noise and $f$ is the Catteneo noise. 
\be
\langle \zeta(\tau)\rangle  = 0 \qquad \langle \zeta(\tau_1)\zeta(\tau_2)\rangle = N(\tau_1) \delta(\tau_1 - \tau_2)   
\ee
It does not matter whether we use $N(\tau_1) $ or $N(\tau_2) $ because of the Dirac-delta function. Let us multiply both sides by the factor $e^{\tau/\tau_Q}$. 
\be 
 \int_{\tau_i}^{\tau}\frac{d}{d\tau} (e^{\tau/\tau_Q} f) d\tau = \int_{\tau_i}^{\tau} \frac{e^{\tau/\tau_Q}}{\tau_Q} \zeta d\tau
 \ee
 \be 
 e^{\tau/\tau_Q} f(\tau)  -  e^{\tau_i/\tau_Q} f(\tau_i) = \frac{1}{\tau_Q} \int^{\tau}_{\tau_i} \zeta(\tau)e^{(\tau'-\tau)/\tau_Q}d\tau'
\ee 
 Let us set $f(\tau_i) = 0$. Another way to see this is in an equilibrium system, the system does not have any initial conditions to be sensitive to. Any fluctuations in $f(\tau)$ will then be solely due to the action of $\zeta(\tau)$. Now we consider two separate times $\tau_1$, $\tau_2$.
\ba
\begin{aligned}
 \langle f(\tau_1)f(\tau_2) \rangle &=\frac{N}{\tau_Q^2} \int_{\tau_i}^{\tau_1} e^{(\tau''-\tau_1)/\tau_Q}d\tau''\int_{\tau_i}^{\tau_2} e^{(\tau'-\tau_2)/\tau_Q}d\tau' \delta(\tau_1 - \tau_2) \\
  &= \frac{N}{\tau_Q^2} \int_{\tau_i}^{\text{min}(\tau_1,\tau_2)} e^{(2\tau'' -\tau_1-\tau_2)/\tau_Q}d\tau''  
 = \frac{N}{2 \tau_Q} \left[ e^{|\tau_1-\tau_2|/\tau_Q} - e^{(2\tau_i -\tau_1-\tau_2)/\tau_Q} \right] 
\end{aligned}
\ea

\section{Constructing the self-correlations}

Self-correlations are defined by Eq.~\eqref{self_correlations}.  As discussed above, their dynamics can be modeled by an equation whose (Fourier transformed) Green's function is related to the original Green's function by
\be
	\tilde G_{\mathrm{self}}(k, \tau, \tau') = \frac{\tilde G(k, \tau, \tau')}{i k}
\ee

The original Green's function is defined schematically by the stochastic differential equation
\be
	D_1 X(\tau, \xi) = D_2 \frac{\partial f}{\partial \xi}(\tau, \xi),
	\label{schematic_EOM_xi}
\ee
where $D_1$ and $D_2$ are differential operators which contain no explicit $\xi$-dependence (other than $\xi$-derivatives) and $f$ is the noisy source.  Fourier transforming the $\xi$-dependence to $k$ as before, this equation becomes
\be
	\tilde D_1 \tilde X(\tau, k) = i k \tilde D_2 \tilde f(\tau, k)
\ee
and its solution is written in terms of the original Green's function as
\be
	\tilde{X}(k, \tau) = -\int_{\tau_0}^\tau d\tau' \tilde{G}(k; \tau, \tau') \tilde{f}(k,\tau')
\ee
We therefore seek an `unphysical' field $X_{\mathrm{self}}$ whose two-point function corresponds to the self-correlations which need to be subtracted out.  This field solution will be generated by the expression
\ba
	\tilde{X}_{\mathrm{self}}(k, \tau)
	&=& -\int_{\tau_0}^\tau d\tau' \tilde{G}_{\mathrm{self}}(k; \tau, \tau') \tilde{f}(k,\tau') \nonumber\\
	&=& -\int_{\tau_0}^\tau d\tau' \frac{\tilde{G}(k; \tau, \tau')}{i k} \tilde{f}(k,\tau') \nonumber\\
	&=& -\int_{\tau_0}^\tau d\tau' \tilde{G}(k; \tau, \tau') \l( \frac{\tilde{f}(k,\tau')}{i k} \r)
\ea
The self-correlations can be therefore straightforwardly generated by replacing $\tilde f \to \tilde f / (i k)$ in $k$-space, which amounts to discarding the $\xi$-derivative in Eq.~\eqref{schematic_EOM_xi}.  Thus, the unphysical self-correlation field is generated by solving the modified equation
\be
	D_1 X(\tau, \xi) = D_2 f(\tau, \xi)
\ee


\end{document}